\documentclass[iop,twocolumn,apj,numberedappendix,appendixfloats]{emulateapj}
\usepackage{apjfonts,color,soul}

\usepackage[pagebackref=false,letterpaper=true,colorlinks=true,citecolor=blue,linkcolor=blue,breaklinks=true,bookmarks=true]{hyperref}

\shorttitle{CGM of SMGs II. ALMA Observations of SMG-QSO Pairs} 
\shortauthors{Fu et al.}
\journalinfo{ApJ accepted}
\submitted{ApJ accepted}

\newcommand{\hers}{{\it Herschel}}

\newcommand{\msun}{$M_{\odot}$}
\newcommand{\msunyr}{$M_{\odot}$\,yr$^{-1}$}
\newcommand{\lsun}{$L_{\odot}$}
\newcommand{\um}{$\mu$m}

\begin{document}

\title{The Circumgalactic Medium of Submillimeter Galaxies. II. Unobscured QSOs within Dusty Starbursts and QSO Sightlines with Impact Parameters below 100 Kiloparsec} 

\author{
Hai~Fu\altaffilmark{1}, Jacob~Isbell\altaffilmark{1}, Caitlin~M.~Casey\altaffilmark{2}, Asantha~Cooray\altaffilmark{3}, J.~Xavier~Prochaska\altaffilmark{4}, Nick~Scoville\altaffilmark{5}, and Alan~Stockton\altaffilmark{6}
}
\altaffiltext{1}{Department of Physics \& Astronomy, University of Iowa, Iowa City, IA 52242}
\altaffiltext{2}{Department of Astronomy, the University of Texas at Austin, 2515 Speedway Blvd, Stop C1400, Austin, TX 78712}
\altaffiltext{3}{Department of Physics and Astronomy, University of California, Irvine, CA 92697}
\altaffiltext{4}{Department of Astronomy and Astrophysics, UCO/Lick Observatory, University of California, 1156 High Street, Santa Cruz, CA 95064}
\altaffiltext{5}{California Institute of Technology, MC 249-17, 1200 East California Boulevard, Pasadena, CA 91125}
\altaffiltext{6}{Institute for Astronomy, University of Hawaii, 2680 Woodlawn Drive, Honolulu, HI 96822}

\begin{abstract}
We present Atacama Large Millimeter/submillimeter Array (ALMA) 870~\um\ observations of 29 bright {\it Herschel} sources near high-redshift QSOs. The observations confirm that 20 of the {\it Herschel} sources are submillimeter-bright galaxies (SMGs) and identify 16 new SMG$-$QSO pairs that are useful to studies of the circumgalactic medium (CGM) of SMGs. Eight out of the 20 SMGs are blends of multiple 870~\um\ sources. The angular separations for six of the {\it Herschel}-QSO pairs are less than 10\arcsec, comparable to the sizes of the {\it Herschel} beam and the ALMA primary beam. We find that four of these six ``pairs'' are actually QSOs hosted by SMGs. No additional submillimeter companions are detected around these QSOs and the rest-frame ultraviolet spectra of the QSOs show no evidence of significant reddening. Black hole accretion and star formation contribute almost equally in bolometric luminosity in these galaxies. The SMGs hosting QSOs show similar source sizes, dust surface densities, and SFR surface densities as other SMGs in the sample. We find that the black holes are growing $\sim$3$\times$ faster than the galaxies when compared to the present-day black-hole-galaxy mass ratio, suggesting a QSO duty cycle of $\lesssim$30\% in SMGs at $z \sim 3$. The remaining two {\it Herschel}-detected QSOs are undetected at 870~\um\ but each has an SMG ``companion'' only 9\arcsec\ and 12\arcsec\ away (71 and 95~kpc at $z = 3$). They could be either merging or projected pairs. If the former, they would represent a rare class of ``wet-dry'' mergers. If the latter, the QSOs would, for the first time, probe the CGM of SMGs at impact parameters below 100\,kpc.  
\end{abstract}


\section{Introduction} \label{sec:intro}

QSOs are among the first high-redshift galaxies that were shown to have luminous thermal (sub)millimeter emission \citep{McMahon94,Isaak94}. Systematic single-dish observations, mostly carried out with the Institut de Radioastronomie Millimetrique (IRAM) 30-m telescope and the James Clerk Maxwell Telescope (JCMT), have found that about 1/3 of optically selected QSOs at $z > 1$ show (sub)millimeter continuum at milliJansky level \citep[e.g.,][]{Omont01,Omont03,Carilli01,Priddey03}. Their brightness is comparable to the Submillimeter-bright Galaxies \citep[SMGs; e.g.,][]{Smail97}, which are JCMT-unresolved sources with 850~\um\ flux densities above $2-3$~mJy. The detection fraction is significantly above that predicted by chance superposition, implying that the QSOs are physically associated with the (sub)millimeter sources. The (sub)millimeter emission indicates rest-frame far-infrared (IR) luminosities on the order of $10^{13}$~\lsun. Detection of molecular gas in these submillimeter-bright QSOs \citep[see][for a review]{Carilli13} indicate that the far-IR luminosities are powered by intense star formation with star formation rates (SFR) of $\sim1,000$~\msunyr. However, the spatial resolution of the single-dish observations at (sub)millimeter wavelengths is limited: e.g., the FWHM beam is 10.6\arcsec\ for the IRAM 30-m at 1.2~mm and 13.8\arcsec\ for the JCMT at 850~\um. Thus, optically-selected QSOs detected in single-dish (sub)millimeter observations may be a mixed population. High-resolution interferometer observations have found examples for the following four categories: (1) SMG-QSO composite galaxy -- QSOs hosted by dusty starbursts \citep[e.g.,][]{Guilloteau99,Wang13,Willott13,Trakhtenbrot17}, (2) gas-rich mergers -- SMG-QSO composite galaxies interacting with nearby dusty starbursts \citep[e.g.,][]{Omont96a,Carilli02,Clements09,Trakhtenbrot17}, (3) gas-rich$-$gas-poor mergers (``wet-dry'' mergers) -- QSOs with gas-poor hosts merging with nearby dusty starbursts \citep[e.g.,SMM\,J04135$+$10277 at $z = 2.84$;][]{Riechers13}, or (4) a line-of-sight projected SMG-QSO pair \citep[e.g., SDSS\,J171209$+$600144 at $z = 2.821$;][]{Fu16}. The last two categories appear to be the least common, with only one example in each category so far. 

\begin{figure*}[!tb]
\epsscale{1.15}
\plotone{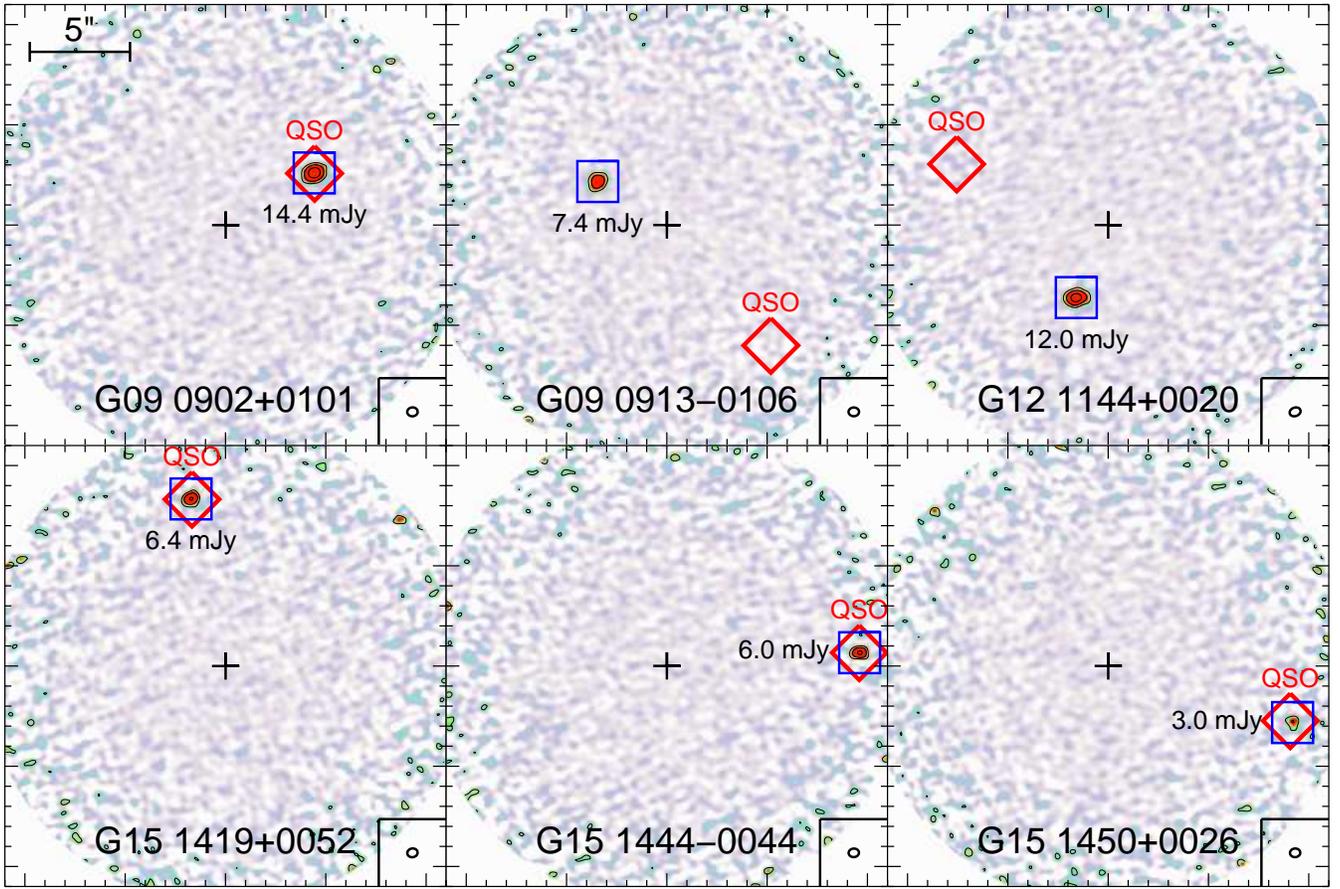}
\caption{ALMA 870~\um\ continuum maps of the \hers-detected QSOs. These are $z \sim 3$ QSOs with angular separations between 5\arcsec\ and 10\arcsec\ from a \hers-selected SMG. The ALMA phase centers were set at the \hers\ 250~\um\ positions ({\it black crosses}). The ALMA 870~\um\ detections are indicated by the blue squares with the integrated 870~\um\ flux densities labeled. The contours are at (0.8, 2, 5)~mJy~beam$^{-1}$. The optical positions of the QSOs are mark by the red diamonds. Whenever the QSO optical position overlaps with an ALMA detection, the two positions differ no more than 0.15\arcsec, which is less than the astrometric uncertainty. The ellipses on the lower right show the shapes of the {\sc clean} beams. ALMA detected a single source in every field and showed that four sources are SMG-QSO composites and two are SMG-QSO pairs. 
\label{fig:alma_qso}} 
\epsscale{1.0}
\end{figure*}

The wide-area far-IR surveys of the {\it Herschel}\footnote{\hers\ is an ESA space observatory with science instruments provided by European-led Principal Investigator consortia and with important participation from NASA.} Space Observatory \citep{Pilbratt10} enabled investigations on the dust-obscured star formation in large samples of optically-selected QSOs from spectroscopic surveys. The 3.5-meter space telescope has a resolution of FWHM = 17.8\arcsec\ and 35.2\arcsec\ at 250~\um\ and 500~\um, respectively. Thus, the \hers\ positions have large uncertainties: the positional offsets between \hers\ sources and their Very Large Array (VLA) 6~GHz counterparts range between 2.7\arcsec\ and 7.9\arcsec\ \citep{Fu16}. Previous studies have used a matching radius of 3-5\arcsec\ to identify the far-IR counterparts of the optically-selected QSOs \citep[e.g.,][]{Cao-Orjales12,Ma15a,Dong16a,Pitchford16}. The conservative matching radii may have resulted in reliable counterpart identification. However, like optically-selected QSOs detected in single-dish (sub)millimeter observations, \hers\ far-IR-detected QSOs are likely a mixed population and only interferometer observations can reveal its constituents. 

The advent of Atacama Large Millimeter/submillimeter Array (ALMA) enables detailed studies of the host galaxies and close environment of high-redshift QSOs. Previous ALMA observations of QSO hosts include dust continuum and [C\,{\sc ii}]\,$\lambda$158~\um\ imaging of six millimeter-detected unobscured QSOs at $z \sim 6$ in band-6 (250~GHz) \citep{Wang13,Willott13}, six QSOs at $z \sim 4.8$ in band-7 (350~GHz) \citep{Trakhtenbrot17}, and CO(3-2) imaging of four heavily reddened QSOs at $z \sim 2.5$ in band-3 (100~GHz) \citep{Banerji17}. These observations have revealed a population of luminous QSOs hosted by dusty starbursts with SFR $\sim$ 1,000 \msunyr, which represent an important co-evolution phase when intense star formation and rapid black hole growth occur simultaneously.

As part of our program to probe the circumgalactic medium (CGM) of SMGs with QSO absorption line spectroscopy, we obtained ALMA band-7 observations for 29 bright \hers\ sources near optically bright QSOs at $z > 2.5$ (i.e., purportedly projected SMG$-$QSO pairs). The positional offset between the optical QSO and the \hers\ 250~\um\ detection ($\theta_{250}$) is required to be between 5\arcsec\ and 30\arcsec. We excluded pairs with $\theta_{250} < 5\arcsec$ because the \hers\ sources are most likely the QSOs. Here we present the ALMA observations; the QSO absorption line study of the ALMA sample will be presented in a future publication. The paper is organized as follows.
In Section~\ref{sec:obs}, we describe the sample selection, the ALMA observations, and the data reduction procedure. 
In Section~\ref{sec:analysis}, we analyze the ALMA images that provide a much shaper view of the sources that are responsible for the \hers\ far-IR emission. In the process, we identified four QSOs hosted within SMGs (i.e., the SMG-QSO composite galaxies).
In Section~\ref{sec:composite}, we present the properties of the SMG-QSO composite galaxies and compare them with other \hers-selected SMGs in the sample.  
We conclude with a summary of our main results in Section~\ref{sec:summary}. 
The Appendix includes the ALMA images and the source catalog.
Throughout we adopt a $\Lambda$CDM cosmology with $\Omega_{\rm m}=0.27$, $\Omega_\Lambda=0.73$ and $H_0$ = 70 km~s$^{-1}$~Mpc$^{-1}$.
\newline
\section{Sample Selection and ALMA Observations} \label{sec:obs}

The \hers\ sources are selected from three equatorial fields of the \hers\ Astrophysical Terahertz Large Area Survey (H-ATLAS) survey \citep{Eales10}. The \hers\ maps cover a total of 161.6~deg$^2$ and overlap substantially with the Galaxy And Mass Assembly \citep[GAMA;][]{Driver16} fields at R.A. = 9, 12, and 15 hr. The average 1$\sigma$ noise of the maps, including both confusion and instrumental noise, is 7.4, 9.4, and 10.2~mJy at 250, 350 and 500~\um\ \citep{Valiante16,Bourne16}. We begin with a sample of \hers-selected SMGs, which are a small subset of \hers\ sources. They are sources with SNR $>$ 3 in all three SPIRE \citep[Spectral and Photometric Imaging Receiver;][]{Griffin10} bands (250, 350, and 500~\um), the SPIRE photometry peaking at 350~\um\ (i.e., ``350~\um\ peakers''), and 500~\um\ flux density $S_{500} > 20$~mJy. We then cross-match the SMGs with optically-selected QSOs from a compilation of spectroscopic surveys and select the SMG$-$QSO pairs with apparent separations between 5\arcsec\ and 30\arcsec. The apparent separation ($\theta_{250}$) is defined as the angular offset between the QSO's optical position and the 250~\um\ position of the {\it nearest} SPIRE source. These {\it apparent} separations are likely different from the {\it true} separations based on ALMA images because of the significant positional uncertainty of SPIRE sources (\S~\ref{sec:posunc}). Refer to \citealt{Fu16} for details about the parent sample. Finally, the ALMA targets in each of the three H-ATLAS/GAMA fields were selected to be located within a $\sim$10~deg diameter circle so that they can be observed in a single scheduling block (SB), optimizing the survey efficiency. 

The ALMA Cycle 3 observations targeting 29 \hers\ SMGs were carried out on 2016 Mar 14 and Mar 30 (Project code: 2015.1.00131.S). A total of 38 to 44 12-m antennae were used, with a minimum baseline of 15.1~m and a maximum baseline of 460~m. This configuration provides a synthesized beam of $\sim$0.5\arcsec\ in FWHM and ensures no flux on scales less than $\theta_{\rm MRS}$ = 7.3\arcsec\ is resolved out by the interferometer (see Table 7.1 in the \href{https://almascience.nrao.edu/documents-and-tools/cycle4/alma-technical-handbook}{ALMA Cycle 4 Technical Handbook}). We used the band 7 (343~GHz/874~\um) receivers with 4 spectral windows of 2~GHz-bandwidth. The spectral windows are centered at 337.5, 339.4, 347.5, and 349.5~GHz. We set the field centers at the \hers\ 250~\um\ catalog positions. The ALMA 12-m antennas' primary beam has a Full Width at Half Power (FWHP) of 17\arcsec\ at 874~\um\ (${\rm FWHP} \simeq 1.13 \lambda/D$; see \S~3.2 in the \href{https://almascience.nrao.edu/documents-and-tools/cycle4/alma-technical-handbook}{ALMA Cycle 4 Technical Handbook}), comparable to the 17.8\arcsec\ FWHM of the \hers\ PSF at 250~\um. This ensures a high detection rate given the uncertainty of the \hers\ positions. 

Each SB lasted between 47 and 51~min, giving a typical on-source integration time of 200~s per target. All science targets within a SB shared the track to optimize the $uv$ coverage. For calibrations in the G09, G12, and G15 fields, we observed respectively (1) the QSOs J0854$+$2006, J1229$+$0203, and J1334$-$1257 for bandpass and pointing calibrations, (2) the QSOs J0909$+$0121, J1150$-$0023, and J1410$+$0203 for phase and amplitude gain calibration, and (3) the QSOs J0854$+$2006, J1229$+$0203, and Titan (\href{https://science.nrao.edu/facilities/alma/aboutALMA/Technology/ALMA_Memo_Series/alma594/abs594}{Butler-JPL-Horizons 2012 models}) for absolute flux calibration. We assume a flux calibration uncertainty of 5\% (see \S~C.4.1 in the \href{https://almascience.nrao.edu/documents-and-tools/cycle4/alma-technical-handbook}{ALMA Cycle 4 Technical Handbook}) and add it in quadrature to the error of the measured ALMA fluxes. 

For imaging, we downloaded the reduced measurement sets from the North American ALMA Science Center (NAASC), which have been fully calibrated by the ALMA pipeline in the Common Astronomical Software Applications (CASA) package \citep[v4.5.1;][]{McMullin07}. We then ran the CASA task {\sc clean} to Fourier transform the calibrated visibilities, to iteratively deconvolve the dirty beam, and to re-convolve with a {\sc clean} beam. We applied the Briggs weighting to the visibilities with a robustness parameter of 0.5 for an optimal balance between sensitivity and spatial resolution. We adopted a {\sc clean} loop gain of 0.1 and restricted the {\sc clean}ing regions with 1\arcsec-radius circular masks around all of the detected sources above 3$\sigma$ level. The {\sc clean} loop continues until the residuals inside the masks reach a threshold of 0.15~mJy/beam. Following common practice, we chose the {\sc clean} beam as the best-fit elliptical Gaussian to the main lobe of the dirty beam. The {\sc clean} beams are on average 0.56\arcsec$\times$0.43\arcsec, adequately sampled by the pixel size of 0.08\arcsec. The {\sc clean}'ed images have rms noises of 0.12 mJy/beam in the G09 and G12 fields and 0.14 mJy/beam in the G15 field. These are consistent with the expectation because only 38 antennae were used in the G15 field while 43/44 antennae were used in the G09/G12 fields. Finally, we construct primary beam corrected images with the CASA task {\sc impbcor} and the primary beam pattern estimated by {\sc clean}. 

\begin{figure}[!t]
\epsscale{1.15}
\plotone{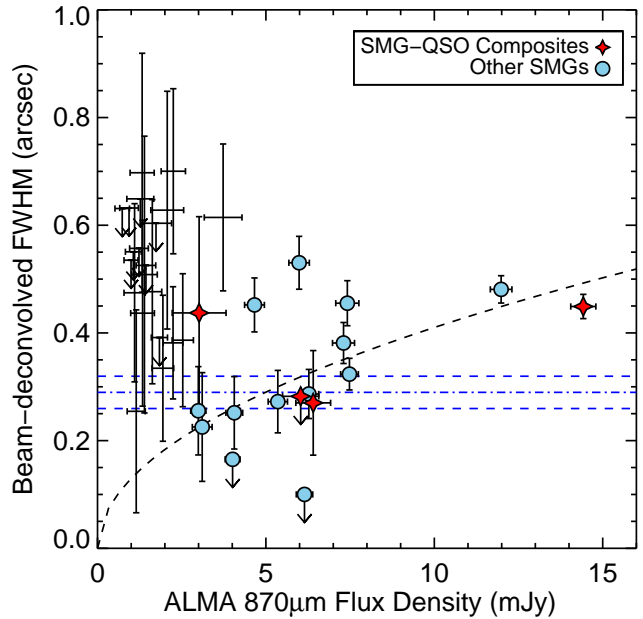}
\caption{Intrinsic sizes of the ALMA sources vs. flux density. We show the four SMG-QSO composite galaxies (see \S~\ref{sec:HQSOs} and Fig.~\ref{fig:alma_qso}) and the remaining ALMA sources as red stars and blue circles, respectively. Note that only the 16 sources with $S_{870}/\delta S_{870} > 10$ (the high-SNR sample) are highlighted with color symbols, with the exception of the SMG-QSO composite G15\,1450$+$0026 (detected at 3.7$\sigma$; the leftmost red star). The blue horizontal lines shows the mean source size of the high-SNR sample and its 1$\sigma$ uncertainty from bootstrap. The black dashed curve shows a fit to the data points assuming FWHM~$\propto \sqrt{S_{870}}$, which is a relation that preserves the star-formation surface density (see \S~\ref{sec:sources} for details). 
\label{fig:size_flux}} 
\end{figure}

\section{Analysis of the ALMA Sources} \label{sec:analysis}

In this section, we focus on the analysis of the ALMA data. We present the measurements of the ALMA counterparts of the \hers\ sources (\S~\ref{sec:sources}), examine the submillimeter emission from the QSOs and their host/companion galaxies (\S~\ref{sec:HQSOs}), and empirically calibrate the formula for \hers\ positional uncertainty (\S~\ref{sec:posunc}).

\subsection{Properties of the ALMA Sources} \label{sec:sources}

For each field, we generate a 22\arcsec$\times$22\arcsec\ {\sc clean}'ed map along with a primary beam pattern. We search for $>$4$\sigma$ peaks\footnote{Here $\sigma$ means the root-mean-square of the {\sc clean}'ed map in units of Jy~beam$^{-1}$.} in the SNR maps (Fig.~\ref{fig:alma_snr}) and fit elliptical Gaussians in the flux density maps (Fig.~\ref{fig:alma_all}) to measure source properties. The ALMA primary beam has a FWHP of $\sim$17\arcsec\ at 870~\um, and the power response declines to almost zero at $\sim$18\arcsec\ off-axis. We thus primarily search for sources within 10\arcsec\ of the field center. Figure~\ref{fig:alma_qso} shows the ALMA images of the six fields where the \hers-QSO separations are less than 10\arcsec, which will be discussed further in the next subsection. Using the CASA task {\sc imfit}, we model all of the sources in a given map simultaneously with elliptical Gaussians to measure their centroid positions, flux densities, and beam-deconvolved sizes. The task also estimates the uncertainties of the fitted parameters following \citet{Condon97}. Because {\sc imfit} cannot handle the spatially variable background noise introduced by the primary beam correction, we ran the task on the {\sc clean}'ed maps without the primary beam correction. The obtained flux densities and their uncertainties are then corrected for the primary beam's power response function. Because of the small source sizes, running {\sc imfit} on the primary beam corrected maps yield almost identical positions and sizes although the errors are significantly underestimated for sources offset from the phase center. 

A total of 39 sources with flux densities between 0.7~mJy~$\leq S_{870} \leq 14.4$~mJy are detected in 27 of the 29 ALMA fields. Table~\ref{tab:alma} in the Appendix presents the source catalog. All of the sources are detected above 5$\sigma$ except one of the two sources in G09\,0918$-$0039 whose peak SNR is 4.8. All of the sources are within 10\arcsec\ of the field center except the 9.6~mJy source in G12\,1132$+$0023, which is 12.3\arcsec\ from the field center. Although we list the source in the catalog, we exclude it in our subsequent analysis so our sample includes 38 ALMA sources. We consider only the fainter 1~mJy source as the counterpart of the \hers\ source in G12\,1132$+$0023. The two fields without ALMA detections are likely spurious \hers\ detections, so they are excluded from our sample. 

These observations confirm that 20 of the 29 (69\%) \hers\ sources in our sample are SMGs with $S_{870} > 2$~mJy. At our resolution, nine of the 29 \hers\ sources break up into multiple 870~\um\ sources. We find that source blending is important even for SMGs fainter than 10~mJy: six of the 16 (38\%) SMGs with $2 < S_{870} < 10$~mJy and two of the four (50\%) SMGs with $S_{870} > 10$~mJy break up into multiples at 0.5\arcsec\ resolution and a noise level of $\sim$0.1~mJy~beam$^{-1}$.

When the SNR is sufficiently high and the beam shape is accurately known, even sources with intrinsic sizes smaller than the beam size can be resolved. Consistent with previous work, here we define the {\it intrinsic source size} as the FWHM of the major axis of the beam-deconvolved Gaussian. Note that previous ALMA studies have shown that the sizes derived from {\sc imfit} are consistent with those derived from directly fitting the $uv$ visibilities with circular Gaussian models \citep{Simpson15,Harrison16}. Previous interferometer observations with the SMA and the ALMA have reported compact, sub-arcsec, sizes of SMGs in strongly lensed SMGs \citep[e.g.,][]{Fu12b,Bussmann13,Bussmann15} and unlensed SMGs \citep[e.g.,][]{Simpson15,Ikarashi15}. Beam-deconvolved source sizes can be reliably measured in the 16 high-SNR ALMA sources, which also have the most reliable flux density measurements ($S_{870}$/$\delta S_{870} > 10$). In Fig.~\ref{fig:size_flux}, we show the ALMA-derived intrinsic angular size against the flux density for all of the ALMA sources, highlighting the 16 high-SNR sources. It is evident that size measurements on sources at lower SNRs are biased to larger values. Thirteen of the 16 high-SNR sources appear resolved, with intrinsic source sizes between $\sim$0.2-0.5\arcsec. The mean source size of those is 0.29\arcsec$\pm$0.03\arcsec, where the uncertainty is from bootstrapping. The mean source size agrees well with that of the ALMA-detected SMGs in the Ultra Deep Survey field \citep[0.30\arcsec$\pm$0.04\arcsec;][]{Simpson15}. 

On the other hand, the two brightest sources ($>$10~mJy) are almost two times larger than the median size ($\sim$0.5\arcsec), suggesting that the source size may increase slowly with flux density. The trend is consistent with a luminosity-independent SFR surface density, which would imply a size-flux-density relation of $R \propto \sqrt{S_{870}}$, given that $S_{870}$ is a good indicator of the IR luminosity (as a result of the negative $K$-correction) and the angular diameter distance decreases by only 8\% between $1.5 < z < 3$. 

\subsection{Submillimeter Emission from the QSOs and their Host/Companion Galaxies} \label{sec:HQSOs}

\begin{figure}[!tb]
\epsscale{1.15}
\plotone{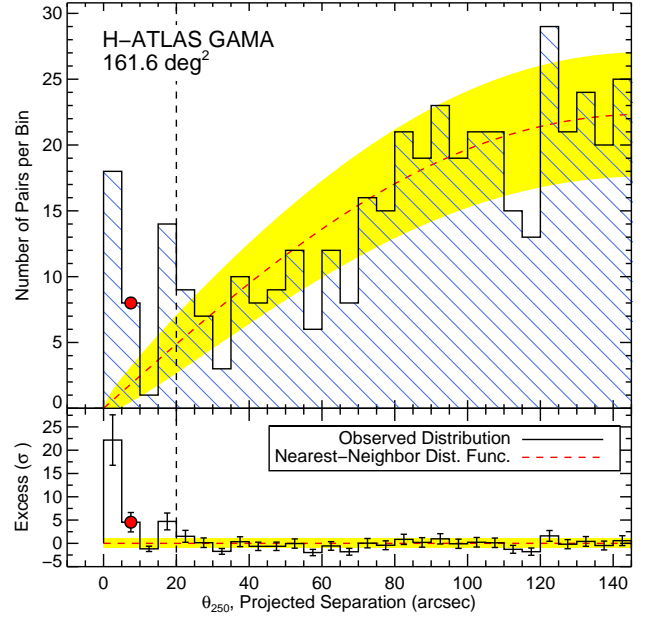}
\caption{Distribution of the angular separation between QSOs and the 250~\um\ positions of \hers-selected SMGs in all three H-ATLAS GAMA fields. The top panel shows a histogram of the observed distribution. The red solid circle highlights the 5-10\arcsec\ bin that includes the six \hers-detected QSOs in Fig.~\ref{fig:alma_qso}. The red dashed curve is the predicted distribution of pairs due to random superpositions, which follows the nearest-neighbor distribution function (Eq.\,\ref{eq:dist}). The yellow stripes delineate the 1$\sigma$ uncertainty of the predicted distribution due to Poisson noise. The bottom panel shows the difference between the observed distribution and the predicted distribution in units of the 1$\sigma$ Poisson noise of the prediction. There is a clear excess over random superpositions at the separation less than 20\arcsec.
\label{fig:sep_dist}} 
\end{figure}

Our ALMA sample can be divided into two parts based on the apparent separation ($\theta_{250}$) between the QSO's optical position and the 250~\um\ position of the nearest \hers-selected SMG. The six pairs with $5\arcsec < \theta_{250} < 10\arcsec$ are the \hers-detected QSOs because the separation is comparable with the FWHM of the \hers\ beam (FWHM = 17.8\arcsec, 24.0\arcsec, and 35.2\arcsec\ at 250, 350, and 500~\um, respectively; \citealt{Valiante16}). More importantly, the QSOs are covered within the sensitive field-of-view of ALMA, given that the FWHP of the ALMA primary beam is $\sim$17\arcsec\ at 874~\um. 

For the remaining 23 fields where $10\arcsec < \theta_{250} < 30\arcsec$, the QSOs are undetected in the ALMA images because (1) they are unlikely far-IR luminous given the nearest \hers\ source is more than 10\arcsec\ away from the QSO position {\it and} (2) the ALMA sensitivity drops severely at off-axis distances beyond 10\arcsec. To provide an upper limit on the intrinsic submillimeter flux density of these QSOs, we stack the ALMA images at the QSO positions for the ten QSOs that are undetected but are within 18\arcsec\ of the field center. Although the rms noise of the stacked image reaches $\sim$0.04~mJy/beam (before primary beam corrections), we did not detect any significant signal at the stacked location. The stacking analysis provides a 3$\sigma$ upper limit of $\sim$3.6~mJy per QSO given the large primary beam correction of $\sim$30$\times$ at the mean off-axis distance of 15\arcsec. It is therefore impossible to detect these QSOs in our ALMA images even if they are bright (mJy-level) sources. 

\begin{figure*}[!t]
\epsscale{0.45}
\plotone{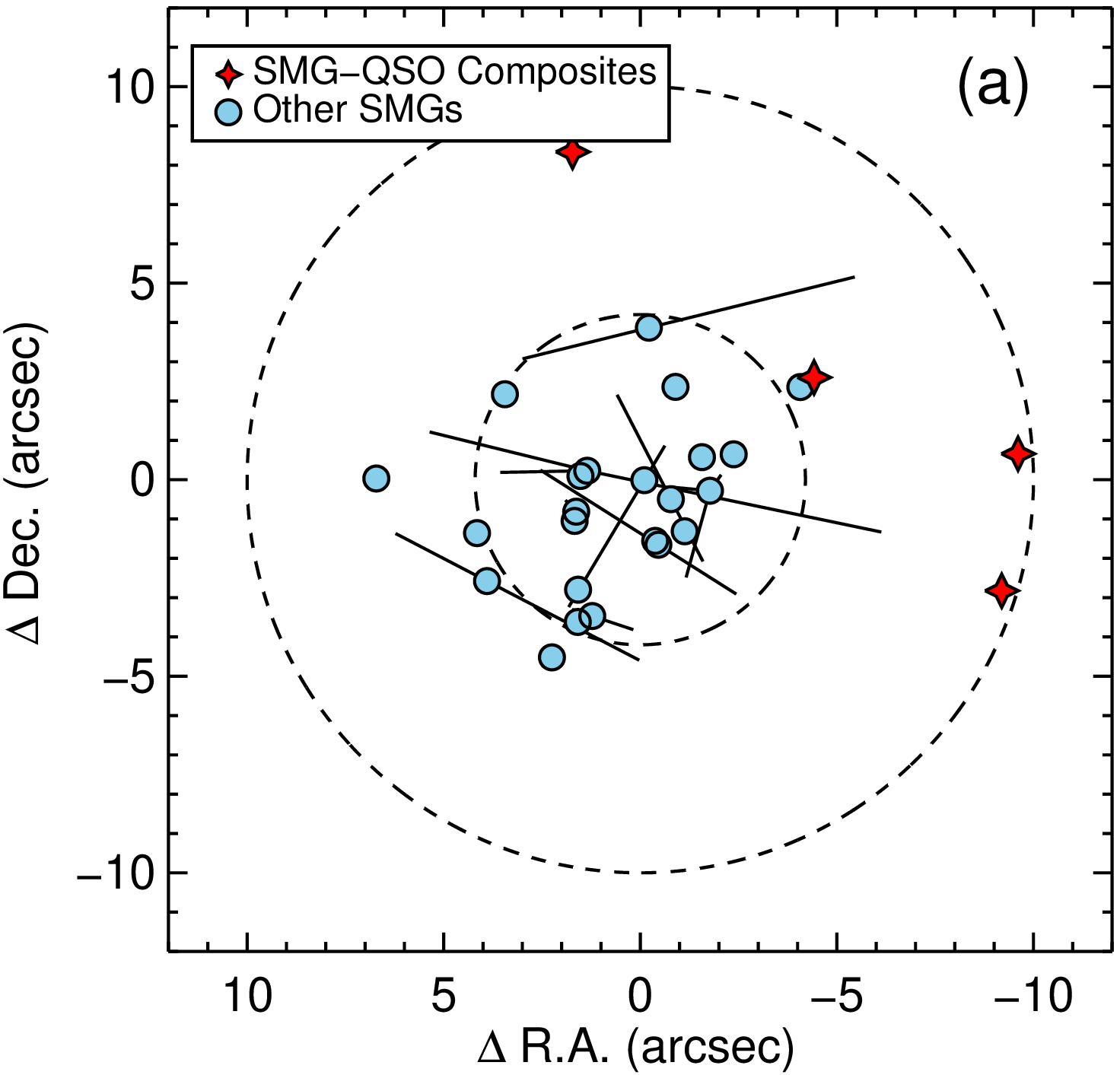}
\plotone{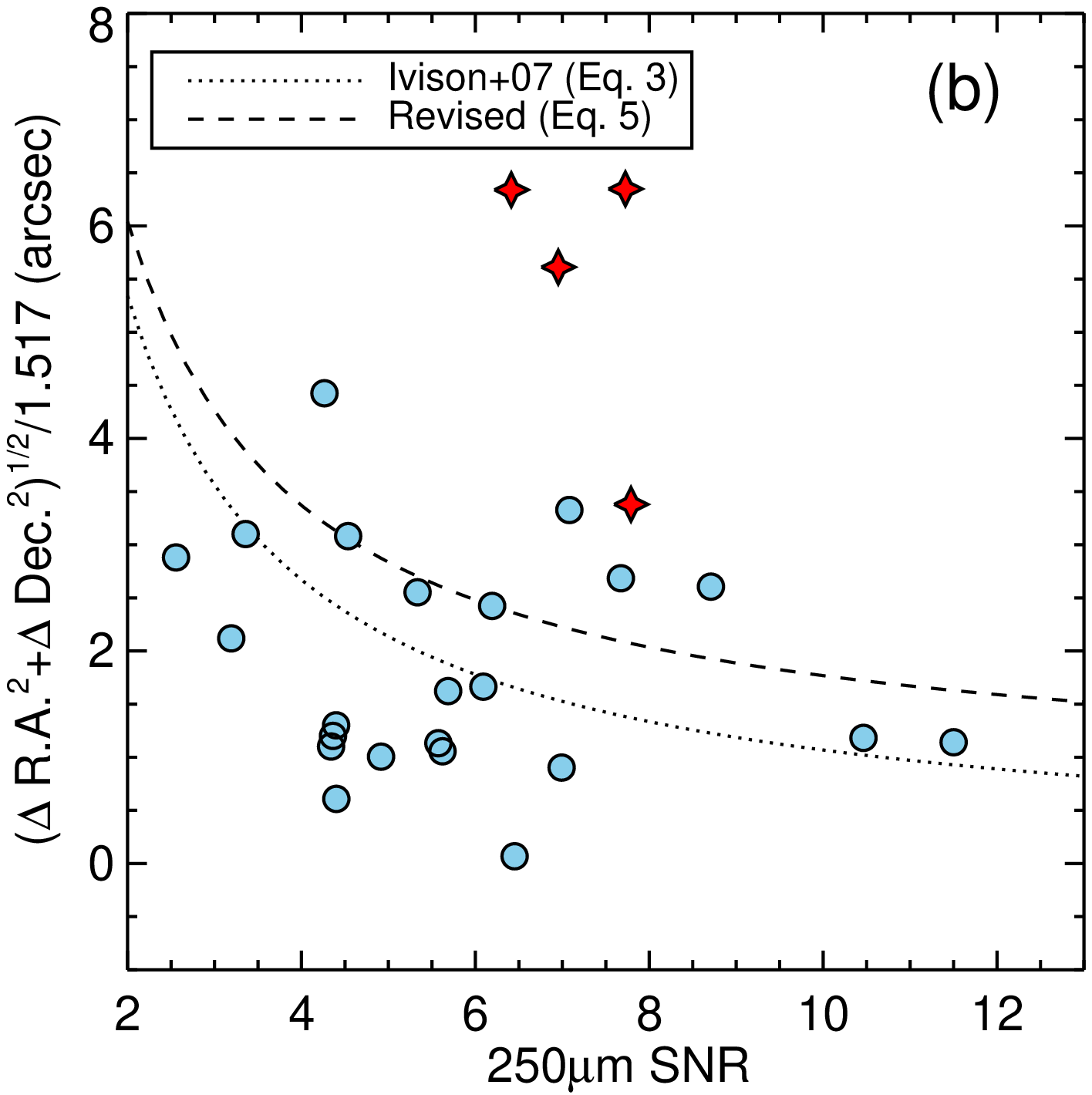}
\caption{Positional offsets between \hers\ 250~\um\ sources and the ALMA counterparts for a total of 27 ALMA fields. For the nine fields that contain multiple ALMA detections, we use the 870~\um\ flux-weighted mean positions. We show the four SMG-QSO composite galaxies and the remaining ALMA sources as red stars and blue circles, respectively. 
({\it a}) Distribution of angular offsets between ALMA positions and \hers\ 250~\um\ positions on the sky. For fields with multiple ALMA detections, we illustrate the extent of the complex with solid lines connecting each component to the mean position. The dashed circles have radii of 4.2\arcsec\ and 10\arcsec, which encloses 67\% (18/27) and 100\% of the sample, respectively.
({\it b}) One-dimensional angular offset vs. the SNR of the \hers\ 250~\um\ detection. We have divided the measured angular offsets by 1.517 so that they can be compared with the theoretical formulae describing the positional uncertainty along one coordinate. Assuming FWHM = 17.8\arcsec\ for the \hers\ PSF, the dotted and dashed curves show the theoretical relation for unresolved sources \citep{Ivison07} and the revised relation (Equation \ref{eq:uncertainty}). The dotted and dashed curves enclose 52\% (14/27) and 67\% (18/27) of the sample, respectively.
\label{fig:offset}} 
\end{figure*}

Figure~\ref{fig:alma_qso} shows the ALMA 870~\um\ continuum images of the six \hers-detected QSOs. In four of the six fields, the QSOs are the only submillimeter sources. The offsets between the ALMA positions and the QSO optical positions ($\theta_{870}$) are less than 0.15\arcsec; and three out of the four sources have $\theta_{870} \lesssim 0.05\arcsec$ (see Table~\ref{tab:qsos}). Such small offsets are comparable to the astrometric uncertainties of our ALMA observations and the SDSS: $\sigma_{\rm 870}^{\rm tot} \simeq {\rm FWHM}/{\rm SNR} = 0.5\arcsec/{\rm SNR}$ \citep{Condon97,Ivison07} and $\sigma_{\rm SDSS}^{\rm tot} \sim 0.1\arcsec$ for point sources brighter than $r \sim 20$~AB \citep{Ivezic02,Pier03}, where the superscript ``tot'' indicates that the uncertainty is measured along two dimensions as opposed to along one coordinate. The 0.15\arcsec\ offset seen in G15\,1450$+$0026 is likely due to the low SNR of the ALMA detection (SNR $\sim$ 3.7). Therefore, we conclude that these offsets are consistent with astrometric uncertainties and these four sources are SMG-QSO composite galaxies, i.e., the QSOs are hosted by dusty starburst galaxies.

In the remaining two fields, the QSOs themselves are undetected by ALMA yet there is a bright ALMA source within 12\arcsec\ of the QSO position. At the redshifts of the two QSOs ($z \sim 2.9$), 10\arcsec\ translates to a transverse proper distance of 80~kpc. Are they interacting pairs or line-of-sight projected pairs? If the former, these cases would represent a rare population of ``wet-dry'' mergers similar to SMM\,J04135$+$10277 at $z = 2.846$ \citep{Riechers13}. They would likely be ``wet-dry'' mergers because the ALMA non-detection of the QSOs implies little interstellar gas in the QSO host galaxies. If the latter, the background QSO sightlines could probe the CGM of SMGs at unprecedentedly small impact parameters (i.e., transverse proper distances). Without spectroscopic redshifts of the submillimeter sources, we attempt to answer this question statistically. 

We can estimate the number of projected pairs in our sample using the surface density of the \hers-selected SMGs ($\Sigma_{\rm SMG}$) and the number of QSOs ($N_{\rm QSO}$) in the GAMA fields. That is because for a Poisson distribution of random positions on the sky, the probability density of finding a nearest neighbor at a distance between $\theta$ and $\theta$+d$\theta$ is described by the nearest-neighbor distribution function \citep{Hertz09}:
\begin{equation}
H(\theta) = \frac{dP}{d\theta} = 2 \pi \Sigma_{\rm SMG} \theta\,{\rm exp}(-\pi \Sigma_{\rm SMG} \theta^2)
\end{equation}
where $\Sigma_{\rm SMG}$ is the average source surface density of the SMGs. Given $N_{\rm QSO}$ random positions sampled by the QSOs in the field, we build up a distribution of $\theta$:
\begin{equation}
\frac{dN}{d\theta} = N_{\rm QSO} \frac{dP}{d\theta} = 2 \pi N_{\rm QSO} \Sigma_{\rm SMG} \theta\,{\rm exp}(-\pi \Sigma_{\rm SMG} \theta^2) \label{eq:dist}
\end{equation}
Given the surface density of \hers-selected SMGs in the three H-ATLAS GAMA fields ($\Sigma_{\rm SMG} = 92$~deg$^{-2}$ over 161.6~deg$^2$) and the number of $z > 2.5$ QSOs in the overlapping area ($N_{\rm QSO} = 1,100$) between the spectroscopic QSO surveys and the \hers\ maps, we overlay the predicted distribution of pairs due to random superpositions in Fig.~\ref{fig:sep_dist} and compare it with the observed distribution of pair separations. The predicted distribution fits the observations nicely at separations greater than 30\arcsec. However, the excess of observed pairs at separations below 30\arcsec\ is evident, indicating an increasing population of physical associations between the SMGs and the QSOs at small angular separations. Note that these physical associations include both QSOs within SMGs and SMG-QSO mergers. In the angular bin of our interest, $5\arcsec < \theta_{250} < 10\arcsec$, there are a total of eight pairs in the GAMA fields. Equation (2) predicts $1.84\pm1.36$ projected pairs in the bin or a fraction of $23\pm19$\%. Since six of these eight pairs were observed by ALMA, we expect only $1.4\pm1.0$ projected pairs (the rest must be physical associations). Given that the ALMA data already prove that four of the six ``pairs'' are submillimeter-bright QSOs, at least one of the remaining two is a projected pair.

\begin{deluxetable*}{ccrcccccccccc}
\tablewidth{0pt}
\tablecaption{Properties of the SMG-QSO composite galaxies.
\label{tab:qsos}}
\tablehead{
\colhead{Object} & \colhead{$z_{\rm QSO}$} & \colhead{$S_{\rm 870}$} & \colhead{offset} & \colhead{FWHM} & \colhead{$L_{\rm IR}^{\rm QSO}$} & \colhead{$L_{\rm IR}^{\rm SF}$} & \colhead{$L_{\rm bol}$} & \colhead{$M_{\rm gas}$} &  \colhead{SFR} & \colhead{$M_{\rm BH}$} & \colhead{$\dot{M}_{\rm BH}$} & \colhead{$\eta_{\rm Edd}$} \\
\colhead{} & \colhead{} & \colhead{mJy} & \colhead{arcsec} & \colhead{kpc} & \colhead{log(L$_\odot$)} & \colhead{log(L$_\odot$)} & \colhead{log(L$_\odot$)} & \colhead{log(M$_\odot$)} & \colhead{M$_\odot$/yr}  & \colhead{log(M$_\odot$)} & \colhead{M$_\odot$/yr} & \colhead{} \\
\colhead{(1)} & \colhead{(2)} & \colhead{(3)} & \colhead{(4)} & \colhead{(5)} & \colhead{(6)} & \colhead{(7)} &  \colhead{(8)} & \colhead{(9)} & \colhead{(10)}  & \colhead{(11)}   & \colhead{(12)}  & \colhead{(13)}
}
\startdata
G09\,0902$+$0101&3.1204& 14.4$\pm$0.8 & 0.03&3.5$\pm$0.2& 12.2$\pm$0.1& 13.2$\pm$0.2& 13.0$\pm$0.2&11.95$\pm$0.12&1700&9.1$\pm$0.4&5.5&0.22\\
G15\,1419$+$0052&2.6711&  6.4$\pm$0.6 & 0.05&2.2$\pm$0.8& 11.8$\pm$0.1& 13.0$\pm$0.2& 12.6$\pm$0.2&11.58$\pm$0.12&1000&9.0$\pm$0.4&2.3&0.12\\
G15\,1444$-$0044&3.3750&  6.0$\pm$0.6 & 0.01&     $<$2.1& 12.2$\pm$0.1& 13.3$\pm$0.2& 13.0$\pm$0.2&11.56$\pm$0.12&1900&9.3$\pm$0.4&6.2&0.18\\
G15\,1450$+$0026&2.8220&  3.0$\pm$0.8 & 0.15&3.5$\pm$1.4& 12.7$\pm$0.1& 12.9$\pm$0.2& 13.1$\pm$0.2&11.04$\pm$0.16& 800&9.1$\pm$0.4&7.1&0.30
\enddata
\tablecomments{
 (1): Designation.
 (2): Spectroscopic redshift of the QSO.
 (3): ALMA 870~\um\ flux density.  
 (4): Offset between the ALMA position and the optical QSO position.
 (5): Beam-deconvolved FWHM along the major axis from elliptical Gaussian fits to the ALMA 870~\um\ image.
(6-7): Rest-frame 8-1000~\um\ luminosity powered by BH accretion and star formation, respectively.
 (8): QSO bolometric luminosity extrapolated from the rest-frame 1350~\AA\ luminosity.
 (9): Total gas mass derived from the AGN-corrected ALMA 870~\um\ photometry.
 (10): SFR from $L_{\rm IR}^{\rm SF}$.
 (11): Virial black hole mass based on C\,{\sc iv}.
 (12): BH accretion rate from the bolometric luminosity of the QSO and a radiative efficiency of $\epsilon = 0.1$.
 (13): Eddington ratio of the QSO.
}
\end{deluxetable*}

\subsection{Herschel Positional Uncertainties} \label{sec:posunc}

Knowing the positional uncertainties is important for identifying the multi-wavelength counterparts of \hers\ sources. With accurate ALMA positions, we can empirically calibrate the expected positional uncertainties. In Fig.~\ref{fig:offset}($a$), we show the spatial offsets between the ALMA positions and the \hers\ 250~\um\ positions. For the nine fields that contain multiple ALMA detections, we use the flux-weighted mean positions. We find that 18 of the 27 fields (67\%) show ALMA positions within 4.22\arcsec$\pm$0.14\arcsec\ of the \hers\ position. This empirical positional uncertainty of $\sigma_{\rm 2D} \simeq 4.2\arcsec$ is appropriate for samples whose median SNR is comparable to that of our sample ($\sim$6.4). 

The theoretical positional uncertainty is a function of the FWHM of the beam and the flux-deboosted SNR \citep{Condon97,Ivison07}: 
\begin{equation}
\sigma_{\rm R.A.} = \sigma_{\rm Dec} = 0.6\times{\rm FWHM}/{\rm SNR}
\label{eq:ivison07}
\end{equation}
In Fig.~\ref{fig:offset}($b$), we show the positional offset between the ALMA and \hers\ positions as a function of \hers\ 250~\um\ SNR, where we have de-boosted the 250~\um\ flux density following \citet{Valiante16}. To directly compare with predictions from the theoretical formula, we have converted the observed angular offsets in two dimension to offsets in one dimension:
\begin{equation}
\Delta_{\rm 1D} = \Delta_{\rm 2D}/1.517 = \sqrt{\Delta_{\rm R.A.}^2 + \Delta_{\rm Dec.}^2}/1.517
\end{equation} 
Note that the radius of the 1$\sigma$ confidence region of a joint two-dimensional normal distribution is 1.517$\times$ larger than that of the corresponding one-dimensional normal distribution. The ratio is not $\sqrt{2}$ as frequently assumed in the literature. Given FWHM = 17.8\arcsec, Eq.~\ref{eq:ivison07} encloses only 14 out of the 27 data points (i.e., 52\% instead of the expected 68\%). This result indicates that the formula have underestimated the true uncertainty for a given SNR, consistent with previous results based on ALMA counterparts of LABOCA sources \citep[e.g.,][]{Hodge13}.

To make the curve enclose 18 out of the 27 fields (i.e., 67\% or 1$\sigma$) in our sample, one can either multiply a factor of 1.4 or add a constant offset of 0.7\arcsec\ to Eq.~\ref{eq:ivison07}. We prefer the latter because the positional uncertainty of \hers\ should not decrease to a fraction of an arcsec even at the highest SNR, as a result of the imperfect \hers\ pointing model and complications introduced in the map making process \citep[e.g.,][]{Smith11}. Therefore, below is our empirically calibrated formula for the positional uncertainty of \hers\ at 250~\um:
\begin{equation}
\sigma_{\rm R.A.} = \sigma_{\rm Dec} = 0.6\times{\rm FWHM}/{\rm SNR} + 0.7\arcsec  \label{eq:uncertainty}
\end{equation}
where FWHM = 17.8\arcsec\ and the SNR is the flux-deboosted SNR. We recommend using Eq.~\ref{eq:uncertainty} to define the search radii of \hers\ counterparts at other wavelengths.

\section{SMG-QSO Composite Galaxies} \label{sec:composite}

In this section, we focus on the physical properties of the four SMG-QSO composite galaxies identified in \S~\ref{sec:HQSOs}. We compare these SMGs that host luminous QSOs with other SMGs in the sample in terms of physical sizes and surface densities of dust mass and SFR in \S~\ref{sec:host}, estimate the black hole virial masses and their accretion rates for the QSOs within SMGs in \S~\ref{sec:BH}, and discuss the implications on the galaxy-BH co-evolution at $z \sim 3$ in \S~\ref{sec:coevol}. The derived properties of the SMG-QSO composite galaxies are tabulated in Table~\ref{tab:qsos}.

\subsection{Properties of the QSO Host Galaxies} \label{sec:host}

We compile and model the spectral energy distributions (SEDs) of the ALMA sources to derive their SFRs and other physical properties. For the SMGs, which may or may not host QSOs, we used photometry from \hers/SPIRE and our ALMA 870~\um\ data. None of our sources are detected by \hers/PACS at 100 and 160~\um, which reach depths of $\sigma = 40$ and 60~mJy, respectively. For the \hers\ sources that are resolved into multiples by ALMA, we use the total 870~\um\ flux densities from all sources so that they are comparable with the \hers\ photometry. For the SMG-QSO composite galaxies, we extend the SEDs with the SDSS photometry between 3500\AA\ and 9000\AA\ \citep{Ahn12} and the SDSS-position ``forced'' Wide-Field Infrared Survey Explorer \citep[WISE;][]{Wright10} photometry between 3.4~\um\ and 22~\um\ \citep{Lang14a}. 

To fix the redshifts of the models, we adopt the QSO spectroscopic redshifts for the four SMG-QSO composite galaxies and assume a redshift of 2.5 for the remaining ALMA sources since they are selected to be ``350~\um\ peakers'' (see \S~\ref{sec:obs}). Optical photometric redshifts are unavailable for the ALMA sources because the depths of the optical and near-IR coverages of the GAMA fields \citep{Driver16} are too shallow to detect the counterparts of the ALMA sources. Note that, due to our sample selection, the SMG-QSO composites are at higher redshifts ($\bar{z}_{\rm QSO} = 3.0$) than the other SMGs ($\bar{z}_{\rm SMG} = 2.5$). 

\begin{figure*}[!tb]
\epsscale{1.15}
\plotone{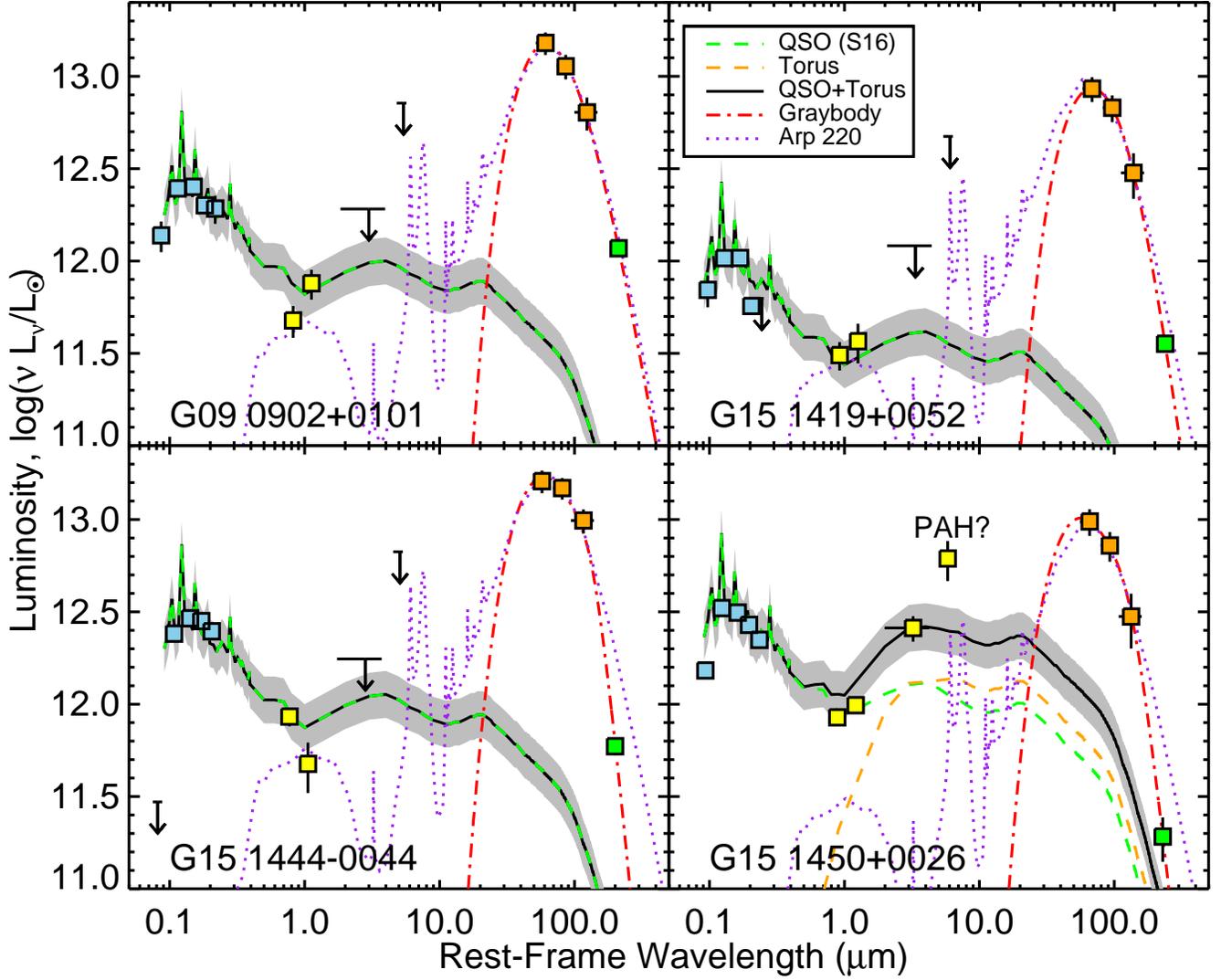}
\caption{Spectral energy distributions of the SMG-QSO composite galaxies. The photometric data points are from the SDSS ({\it blue}), WISE ({\it yellow}), \hers/SPIRE ({\it orange}), and ALMA ({\it green}). We fit the SED at $\lambda_{\rm rest} < 10$~\um\ with a combination of the mean SED of unobscured QSOs \citep[({\it green dashed lines})][]{Symeonidis16} and a ``torus'' component (i.e., the QSO template reddened by $A_V = 4$~mag; {\it orange dashed lines}). The black curve and the gray shaded area show the combined QSO$+$torus SED and the 1$\sigma$ uncertainty (0.13\,dex) of the QSO template. Note that strong PAH emission at 6.3~\um\ may be needed to explain the WISE 22~\um\ photometry of G15\,1450$+$0026. The far-IR excess above QSO emission is evident, which indicates intense star formation in the host/companion galaxies. We overlay the best-fit modified blackbody emission (Eq.~\ref{eq:graybody}) from dusts at $T \sim 55-65~K$ ({\it red dash-dot lines}) and a scaled Arp~220 template \citep[{\it dotted purple lines};][]{Silva98}. 
\label{fig:sed}} 
\end{figure*}

\subsubsection{QSO Contribution to the IR Emission} \label{sec:qso_correct}

\begin{figure*}[!t]
\epsscale{0.45}
\plotone{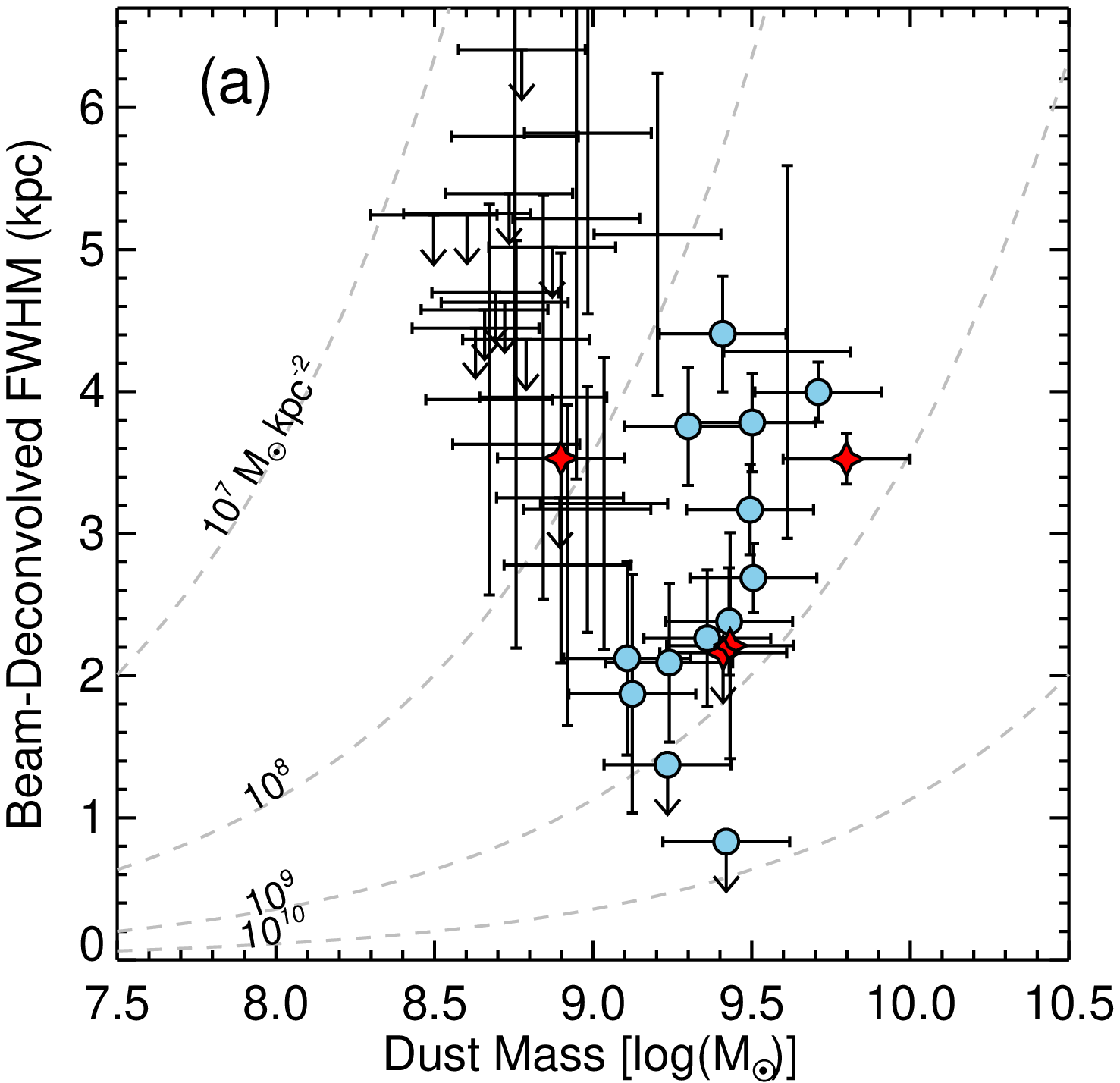}
\plotone{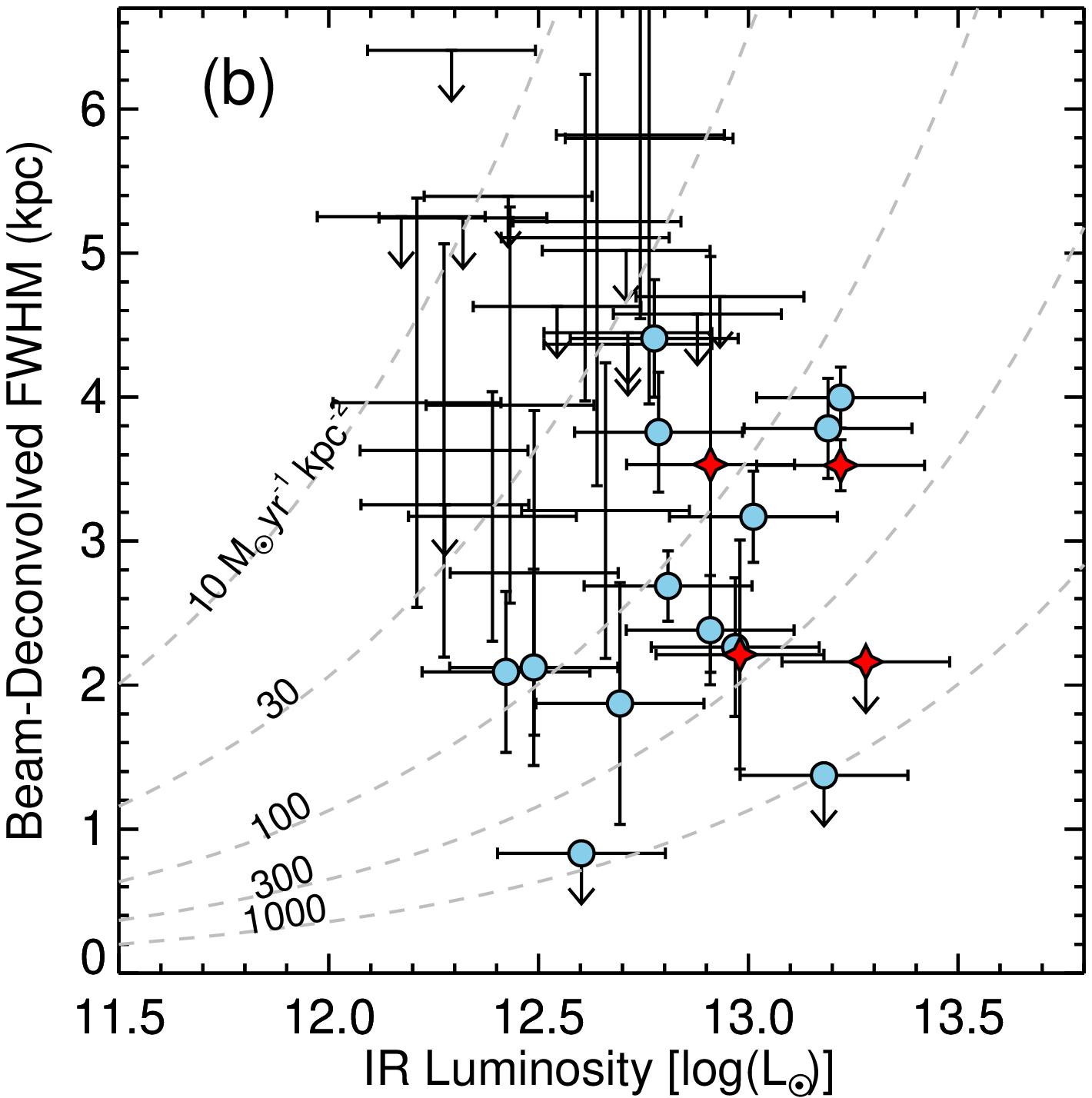}
\caption{Intrinsic submillimeter size vs. dust mass and IR luminosity. Symbols are the same as in Fig.~\ref{fig:size_flux}.
({\it a}) Beam-deconvolved FWHM vs. dust mass. The dust masses are derived from the 870~\um\ flux densities. For the SMG-QSO composites ({\it red diamonds}), the observed 870~\um\ flux densities have been corrected for the QSO contribution. Dashed curves show constant dust surface mass densities for circular areas with radii equal to half of the beam-deconvolved FWHM. The contour levels are labeled.
({\it b}) Beam-deconvolved FWHM vs. IR luminosity. Dashed curves show constant SFR surface densities. For the SMG-QSO composites, we use the QSO-corrected IR luminosities.
\label{fig:size}} 
\end{figure*}

For the SMG-QSO composite galaxies, the far-IR emission could be powered by both star formation and BH accretion. We thus need to correct the observed far-IR and submillimeter photometry for the QSO contribution. Fig.~\ref{fig:sed} shows the observed SEDs of the SMG-QSO composite galaxies. By assuming negligible contribution from dust-obscured star formation at $\lambda_{\rm rest} < 10$~\um, we can fit the SDSS and WISE photometry with a QSO SED template and then estimate the QSO correction in the far-IR by extrapolating the best-fit QSO template. We adopt the intrinsic QSO SED from \citet{Symeonidis16} and extend it below 0.4~\um\ with the composite median QSO spectrum from the SDSS \citep{Vanden-Berk01}. The template is derived from a sample of unobscured luminous PG QSOs at $z < 0.18$. We choose this SED template because it contains a dust component at a cooler temperature than that of \citet{Richards06}. Whether this cooler dust component is due to AGN heating or star formation remains under debate \citep{Lyu17,Lani17}. We thus note that the \citet{Symeonidis16} AGN SED provides a conservative upper bound on the QSO's IR luminosity. 
 
As shown in Fig.~\ref{fig:sed}, the QSO template fits the SDSS and WISE photometry well.  An exception might be G15\,1450$+$0026, which shows extra amount of emission above rest-frame $\sim$3~\um. It is also the only object that is detected by WISE at 12~\um\ and 22~\um. So for this object, we include a separate ``torus'' component to improve the fit \citep{Polletta06}. The torus model represents a dust-obscured QSO component and it is constructed by reddening the QSO template by $A_V = 4$\,mag with the \citet{Calzetti00} extinction law. 

In all four cases, the best-fit QSO($+$torus) models severely under-predict the far-IR emission, indicating that most of the far-IR emission is powered by star formation instead of BH accretion. Puzzlingly, none of the QSOs show evidence of significant dust reddening in the rest-frame UV ($\lambda_{\rm rest} < 0.3$~\um), even though the compact host galaxies appear extremely dusty. Based on the best-fit QSO($+$torus) model, we estimate the QSO contribution to the observed 870~\um\ flux density and the total integrated IR luminosity. The uncertainty of these estimates is at least 0.13~dex, which is the dispersion of individual QSO SEDs around the mean SED.

\subsubsection{Mass and Surface Density of the Interstellar Medium} \label{sec:gasmass}

Following the method of \citet{Scoville16} to estimate the dust mass from the optically thin Rayleigh-Jeans dust emission, we extrapolate the {\it observed} 870~\um\ luminosity to {\it rest-frame} 850~\um\ luminosity using a modified blackbody with $T = 25$\,K, $\beta = 1.8$, and $\lambda_0 = 50$~\um. This procedure is equivalent to using their $K$-correction formula. Note that, for the SMG-QSO composite galaxies, we use the QSO-corrected 870~\um\ flux densities from \S~\ref{sec:qso_correct}. We adopt the mean luminosity-to-gas-mass ratio of $L_\nu^{850}/M_{\rm gas} = (6.7\pm1.7)\times10^{19}$\,erg\,s$^{-1}$\,Hz$^{-1}$\,$M_\odot^{-1}$ and the Milky Way gas-to-dust ratio of $M_{\rm gas}/M_{\rm dust} = 140$ \citep{Draine07} to obtain the dust mass.

To estimate the dust mass surface densities, we plot the beam-deconvolved physical sizes against the dust mass in Fig.~\ref{fig:size}($a$). Note that the size estimates are biased high at ${\rm SNR} < 10$, so these data points should be considered as upper limits. All of the sources detected at sufficiently high SNR ($> 10$) show similar dust mass surface densities: the average dust mass surface density is $\sim8\times10^8$\,\msun\,kpc$^{-2}$ and almost all data points are bracketed between $10^8$ and $10^9$\,\msun\,kpc$^{-2}$. There is apparently no difference between the SMGs hosting luminous QSOs and the other SMGs in the sample, in terms of both the surface density and the total gas/dust mass.

Note that the Rayleigh-Jeans method assumes the Galactic CO-to-H$_2$ conversion factor, $X_{\rm CO} \equiv N_{\rm H_2}/W_{\rm CO} = 3\times10^{20}$\,cm$^{-2}$\,(K\,km\,s$^{-1}$)$^{-1}$ where $W_{\rm CO}$ is the CO line intensity or equivalently $\alpha_{\rm CO} \equiv L^\prime_{\rm CO} / M_{\rm gas} = 6.5$\,\msun\,pc$^{-2}$\,(K\,km\,s$^{-1}$)$^{-1}$ after correcting for Helium and heavier elements ($M_{\rm gas}/M_{\rm H_2} = 1.36$). We thus could have overestimated the dust masses by $4.6\times$ if a lower conversion factor is more suitable for SMGs \citep[e.g., $\alpha_{\rm CO} \simeq 1.4$;][]{Magdis11,Hodge12,Magnelli12b}, increasing the luminosity-to-gas-mass ratio ($L_\nu^{850}/M_{\rm gas}$). Adopting $\alpha_{\rm CO} = 1.4$, the mean dust mass surface density of our sample becomes $\sim2\times10^8$\,\msun\,kpc$^{-2}$.

\subsubsection{IR Luminosity and SFR Surface Density} \label{sec:lir}

We estimate the IR luminosity integrated between $8 < \lambda_{\rm rest} < 1000$~\um\ ($L_{\rm IR}$) by fitting a modified blackbody curve to the \hers/SPIRE and ALMA photometry. We adopt the general solution of the radiative transfer equation assuming local thermal equilibrium at a constant temperature $T$: 
\begin{equation}
S_\nu \propto (1-e^{-\tau_\nu})~B_\nu(T) = (1-e^{-(\lambda/\lambda_0)^{-\beta}})~B_\nu(T)
\label{eq:graybody}
\end{equation}
where $B_\nu(T)$ is the Planck function, $\tau_\nu$ the frequency-dependent optical depth, and $\lambda_0$ is the wavelength at which $\tau_\nu = 1$. We have assumed that the dust opacity follows a power-law with a slope of $\beta$ at wavelengths greater than the dust size (i.e., $\lambda \gtrsim 10$~\um); i.e., $\tau_\nu =  (\nu/\nu_0)^\beta = (\lambda/\lambda_0)^{-\beta}$ and $\beta > 0$. Given the dust mass-absorption coefficient of $\kappa = 0.07$\,m$^2$~kg$^{-1}$ at 850~\um\ for Galactic dust \citep{Dunne00,James02}, it can be shown that $\lambda_0$ depends on the dust surface density:
\begin{equation}
\lambda_0 = 850~{\rm \mu m}~(\Sigma_{\rm dust}/6.8\times10^9 M_\odot~{\rm kpc}^{-2})^{1/\beta}
\end{equation}
where $\Sigma_{\rm dust}$ is the dust surface density. We fixed $\lambda_0$ at 150~\um\ for a typical power-law slope of $\beta = 2$ and a typical dust mass surface density of $\sim2\times10^8$\,$M_\odot~{\rm kpc}^{-2}$. The best-fit dust temperatures range between 35 and 65~K. We note that $L_{\rm IR}$ is insensitive to our choice of $\lambda_0$ and the analytical form of the modified blackbody, because the peak of the dust emission is relatively well constrained by the \hers\ data and the redshift is fixed. Even optically thin models, i.e., $S_\nu \propto \nu^\beta~B_\nu(T)$, give similar $L_{\rm IR}$ despite significantly lower best-fit dust temperatures (20-45~K). 

The 27 \hers\ sources with ALMA detections have IR luminosities between $12.6 < {\rm log}(L_{\rm IR}) < 13.3$~\lsun\ with a mean at 12.9~\lsun. For the SMG-QSO composite galaxies, we subtract the QSO IR luminosity based on the best-fit QSO(+Torus) SED from \S~\ref{sec:qso_correct}. The QSO-corrected IR luminosity, $L_{\rm IR}^{\rm SF} = L_{\rm IR} - L_{\rm IR}^{\rm QSO}$, has an uncertainty of $\sim$0.2~dex. The uncertainty has three components. First, we find that the statistical error of $L_{\rm IR}$ is $\sim$0.1\,dex for sources with spectroscopic redshifts, through Monte Carlo simulations of synthetic \hers\ and ALMA photometry. The systematic uncertainty due to the choice of SED model introduces an additional 0.1~dex uncertainty. For example, in Fig.~\ref{fig:sed} we also show the best-fit model to the far-IR photometry using the SED of the local ultra-luminous IR galaxy (ULIRG) Arp\,220 \citep{Silva98}. The resulting $L_{\rm IR}$ can be higher than the blackbody fits by $\sim$0.1~dex because the Arp\,220 SED contains hotter dusts that gives a shallower Wien's tail. Finally, the extrapolated IR luminosity of the QSO has a minimum uncertainty of 0.13~dex. Combining the three uncertainties in quadrature, we estimate an uncertainty of 0.2~dex for $L_{\rm IR}^{\rm SF}$. 

To estimate the SFR surface densities, we plot the beam-deconvolved FWHMs against the IR luminosity in Fig.~\ref{fig:size}($b$). SFRs are estimated with the \citet{Kennicutt98} calibration for the \citet{Chabrier03} IMF: 
\begin{equation}
{\rm SFR}/M_{\odot}\,{\rm yr}^{-1} =  L_{\rm IR}^{\rm SF}/10^{10}\,L_{\odot}
\label{eq:SFR}
\end{equation}
In Fig.~\ref{fig:size}($b$), we compare our measurements against curves at fixed SFR surface densities. The mean SFR surface density of the SMG-QSO composites is $\sim$260\,\msunyr\,kpc$^{-2}$. G15\,1444$-$0044 shows the highest surface density, its unresolved size places a 3$\sigma$ lower limit at $>$520\,\msunyr\,kpc$^{-2}$, approaching the Eddington limit of $10^3$\,\msunyr\,kpc$^{-2}$ for maximum starbursts \citep{Thompson05}. The results for the rest of the sample are less reliable because of their unknown redshifts, but their mean surface density of $\sim$240\,\msunyr\,kpc$^{-2}$ is almost the same as that of the SMG-QSO composites, and most of the data points are constrained within a narrow surface density range between 100 and 300\,\msunyr\,kpc$^{-2}$. Note that the higher average redshift of the SMG-QSO composite galaxies compared to the other SMGs probably have caused the apparent horizontal offset between the two subsamples. These SFR surface densities are similar to previous ALMA-observed SMGs \citep[e.g.,][]{Ikarashi15,Simpson15,Harrison16}, although only photometric redshifts were used in these previous studies.

\subsection{Properties of the Supermassive Black Holes}\label{sec:BH}
  
We estimate the black hole virial masses with the calibration of \citet{Vestergaard06} that involves the C\,{\sc iv}$\lambda$1500 line width (FWHM$_{\rm CIV}$) and the continuum luminosity at $\lambda_{\rm rest} = 1350$~\AA\ ($L_{1350}$):
\begin{equation}
{\rm log}(\frac{M_{\rm BH}}{M_\odot}) = 6.66 + 0.53\,{\rm log}(\frac{L_{1350}}{10^{44} {\rm erg/s}}) + 2\,{\rm log}(\frac{{\rm FWHM_{CIV}}}{10^3 {\rm km/s}})
\end{equation}
The 1$\sigma$ uncertainty of the zero point is 0.36\,dex. We measure the C\,{\sc iv}$\lambda$1500 line width from the optical spectrum. G15\,1450$+$0026 does not have a SDSS/BOSS spectrum, so we measure the C\,{\sc iv} FWHM using the spectrum from the 2dF QSO Redshift Survey \citep[2QZ;][]{Croom04}. The other five QSOs all have BOSS spectra, so we adopt the line widths in the SDSS Data Release 12 Quasar catalog \citep[DR12Q;][]{Paris16}, which are based on principle component fits. We estimate the QSO continuum luminosity at rest-frame 1350~\AA\ by interpolating the SDSS broad-band photometry. The BH masses are similar among the four QSOs: ${\rm log}(M_{\rm BH}/M_\odot) \simeq 9.1$. 

The broad C\,{\sc iv} lines often show evidence of non-virial kinematics (e.g., outflows), the C\,{\sc iv}-based virial masses are thus biased high when compared with virial masses from Hydrogen Balmer lines \citep[e.g.,][]{Coatman17}. However, we do not correct our BH masses, because the systematic bias is low at $\sim10^{9}$\,\msun\ and the precise systematic redshifts of the QSO host galaxies are currently undefined.

To estimate the BH accretion rates and the Eddington ratios, we apply a bolometric correction to the quasar continuum luminosity with the luminosity-dependent QSO SEDs from \citet{Hopkins07}:
\begin{equation}
{\rm log}(L_{\rm bol}/{\rm erg\,s}^{-1}) \simeq 1.40 + 0.93\,{\rm log}(L_{1350}/{\rm erg\,s}^{-1})
\end{equation}
The bolometric correction has an uncertainty of $\sim$0.2~dex given the level of dispersion among individual QSO SEDs. We find that the bolometric luminosities of the QSOs are comparable to the integrated IR luminosities from star-formation ($L_{\rm IR}^{\rm SF}$). BH accretion contributes a large fraction (30-60\%) of the total bolometric luminosity ($L_{\rm bol} + L_{\rm IR}^{\rm SF}$) in the SMG-QSO composite galaxies, similar to previously studied far-IR-bright QSOs at $z \sim 4.8$ \citep{Netzer14,Trakhtenbrot17}.

Given that $L_{\rm bol} = \epsilon \dot{M}_{\rm BH} c^2 /(1-\epsilon)$, we then assume a radiative efficiency of $\epsilon = 0.1$ to convert the bolometric luminosity to an accretion rate:
\begin{equation}
\dot{M}_{\rm BH}/M_\odot\,{\rm yr}^{-1} = L_{\rm bol}/6.3\times10^{45}\,{\rm erg\,s}^{-1}
\label{eq:BHAR}
\end{equation}
The estimated BH accretion rates range between 2 and 7\,\msun~yr$^{-1}$, with Eddington ratios between 10 and 30\% when compared to the C\,{\sc iv} BH masses (Table~\ref{tab:qsos}). In the next subsection, we discuss the implications of these measurements in the context of co-evolution.

\subsection{Star-Formation-AGN Co-evolution at $z \sim 3$} \label{sec:coevol}

Star formation and supermassive BH accretion seem to be strongly connected, at least when averaged over the $\sim$100~Myr star-formation timescales \citep{Hickox14}. The star-formation-AGN co-evolution scenario is supported by two major pieces of observational evidence. Firstly, the comoving luminosity densities of AGN and star-forming galaxies have shown that the cosmic BH accretion history closely follows the star-formation history over a wide redshift range \citep[e.g.,][]{Boyle98,Hopkins07,Zheng09}. Secondly, the average BH accretion rate of star-forming galaxies scales almost linearly with the SFR \citep[e.g.,][]{Rafferty11,Chen13}.

However, there are some quantitative disagreement between the two. The present-day ratio of the integrated BH mass density \citep[${\rm log}(\rho_{\rm BH}/M_\odot\,{\rm Mpc}^{-3}) = 5.66\pm0.15$;][]{Marconi04} and stellar mass density \citep[${\rm log}(\rho_{\rm star}/M_\odot\,{\rm Mpc}^{-3}) = 8.4\pm0.1$;][]{Bell03} is ${\rm log}(\rho_{\rm star}/\rho_{\rm BH}) = 2.74\pm0.18$. It implies a growth ratio of ${\rm SFR}/\dot{M}_{\rm BH} = 2\times\rho_{\rm star}/\rho_{\rm BH} = 1100^{+560}_{-370}$, provided that 50\% of the formed stellar mass is recycled into the interstellar medium. This is consistent with the ratio of the integrated BH accretion rate ($\dot{M}_{\rm BH}$) and SFR, $\dot{M}_{\rm BH}(z)/{\rm SFR}(z) \simeq (5-8)\times10^{-4}$, based on observed luminosity functions \citep{Hopkins07,Zheng09}.
On the other hand, the average X-ray AGN luminosity for star-forming galaxies implies a growth rate ratio of ${\rm SFR}/\dot{M}_{\rm BH} \simeq 3000$. This is $\sim3\times$ higher than the growth ratios based on the luminosity functions and the present-day mass ratio. 

The discrepancy can be explained if two third of BH accretion occur in AGNs that are intrinsically X-ray under-luminous (thus require larger bolometric corrections). If so, one would expect a higher $\dot{M}_{\rm BH}/{\rm SFR}$ ratio when mid-IR selected AGNs are included. This is indeed the case. Using {\it Spitzer} mid-IR spectra of a complete sample of 24~\um\ selected galaxies at $z \sim 0.7$ in the COSMOS field, we have determined an average AGN bolometric luminosity of $L_{\rm AGN} = 10^{45}$\,erg\,s$^{-1}$ for star-forming galaxies with an average IR luminosity of $L_{\rm IR} = 10^{12}$\,\lsun\ \citep{Fu10}. We estimate a $\sim$0.13~dex bootstrap uncertainty for the mean AGN luminosity, although the uncertainty in the bolometric correction dominates the error budget ($\sim$0.2~dex). The luminosity ratio indicates a growth ratio of ${\rm SFR}/\dot{M}_{\rm BH} = 570\pm230$.

\begin{figure}[!t]
\epsscale{1.1}
\plotone{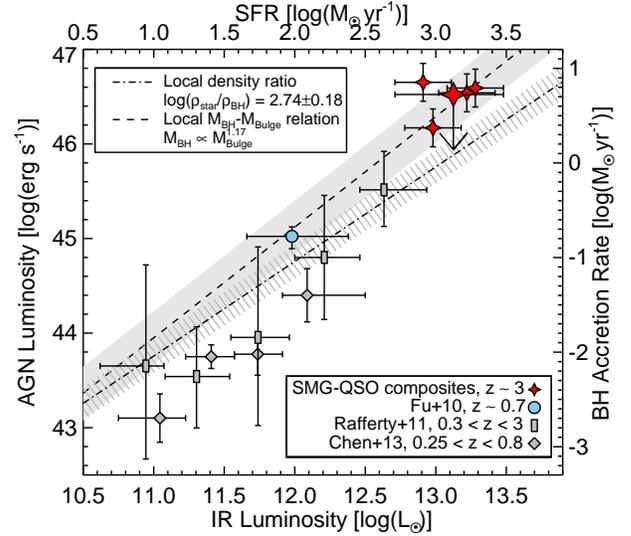}
\caption{AGN bolometric luminosity vs. host galaxy IR luminosity. The corresponding BH accretion rate and the SFR are indicated on the right and top axes, respectively. The SMG-QSO composite galaxies are shown as red diamonds and the QSO contribution has been removed from the IR luminosity. The mean luminosities traced by these data points ({\it big red diamond}) should be treated as an upper limit because the sample do not include SMGs without AGN activity. For example, if $\sim$30\% of the SMGs with $L_{\rm IR} \sim 10^{13}$\,\lsun\ host luminous QSOs, the mean data point would shift downward by the length of the downward arrow (0.5\,dex).
For comparison, we show the average AGN bolometric luminosities vs. the average IR luminosities for samples of star-forming galaxies. Note the different AGN tracers were used to estimate the bolometric luminosity: for our SMG-QSO composites we used the luminosity at 1350~\AA, \citet{Rafferty11} and \citet{Chen13} used 2-10~keV X-ray luminosity, and \citet{Fu10} used 15~\um\ continuum luminosity. For the literature values, the error bars of the X-axis show the range of IR luminosities covered by each subsample, while the error bars of the Y-axis show the bootstrap 1$\sigma$ uncertainty of the mean AGN luminosities. 
The black dot-dashed line and the black dashed line show the growth rates required by the present-day ratio of integrated BH mass density and stellar mass density and the local $M_{\rm BH}-M_{\rm Bulge}$ correlation \citep{Kormendy13}, respectively. The striped and the gray-shaded regions around the lines show the 1$\sigma$ uncertainties of the local relations. See \S~\ref{sec:coevol} for the assumptions used in converting these mass ratios to the luminosity ratios. 
\label{fig:coevol}} 
\end{figure}

Figure~\ref{fig:coevol} compares the requirements on growth rates from the present-day BH-stellar mass ratios and the direct measurements at higher redshifts. It plots the AGN bolometric luminosity against the IR luminosities. These luminosities directly trace the BH accretion rates (Eq~\ref{eq:BHAR}) and the SFRs (Eq.~\ref{eq:SFR}), as shown on the right and top axes of the figure. In a strong co-evolution scenario, the BH accretion rate should scale with the SFR following the present-day BH-galaxy mass ratio at all redshifts. The dash-dotted line and the associated striped region shows the relation required by the present-day mass ratio, i.e., ${\rm SFR}/\dot{M}_{\rm BH} = 1100^{+560}_{-370}$. For comparison, the dashed line and the associated gray-shaded region shows the local $M_{\rm BH}-M_{\rm bulge}$ correlation and its 1$\sigma$ scatter of 0.28\,dex \citep{Kormendy13}. Because it is a non-linear relation, we make the following assumption to convert the mass ratio to the growth rate ratio: (1) a 50\% gas fraction ($M_{\rm gas} = 0.5\,M_{\rm Bulge}$), (2) a gas exhaustion timescale of $\tau = 2\,M_{\rm gas}/{\rm SFR} = 200$\,Myr, and (3) a 50\% recycling rate of the stellar mass. Note that the location of the curve is not sensitive to the detail values of these parameters. Because the bulge mass is a fraction of the total stellar mass, the relation requires lower SFRs at any given BH accretion rate. The dashed line is thus systematically shifted to the left side of the dash-dotted line.

Then, we show the average AGN luminosities for star-forming galaxies at various IR luminosities from \citet{Fu10}, \citet{Rafferty11}, and \citet{Chen13}. Note that \citet{Fu10} performed spectral decomposition on the mid-IR spectra and used the 15~\um\ luminosity from the AGN to extrapolate the bolometric luminosities, while \citet{Rafferty11} and \citet{Chen13} extrapolated 2-10~keV X-ray luminosity to obtain the AGN luminosities. For these literature values, the error bars of the X-axis show the range of IR luminosities covered by each subsample, while the error bars of the Y-axis show the bootstrap 1$\sigma$ uncertainty of the mean QSO luminosities. As described above, many of the X-ray-derived AGN luminosities from \citet{Rafferty11} and \citet{Chen13} are systematically lower (by up to $\sim$0.5~dex) than that required to explain the present-day mass ratio, while the mid-IR-derived average AGN luminosity from \citet{Fu10} is consistent with that required by the present-day BH-stellar density ratio within the uncertainties.

Finally, we plot the SMG-QSO composite galaxies from this study. We have subtracted the QSO contribution to the IR luminosities so that they trace star formation only. These composite galaxies offer us a glimpse into the co-evolution phase of massive galaxies at $z \sim 3$ in an extremely luminous regime that has not been probed in previous studies. As QSOs selected to be IR-luminous, this sample do not include SMGs without AGN activity. Hence, the average QSO luminosity of our sample provides an upper limit on the average QSO luminosity of IR-selected galaxies with $L_{\rm IR} \sim 10^{13}$\,\lsun.  A rough estimate of the AGN fraction in SMGs can be made using the offset between the observed QSO luminosity and the relation based on the present-day density ratio. If we were to shift the average QSO luminosity downward by 0.5 dex so that it reaches the level required by the present-day density ratio, then it implies a QSO duty cycle of $\sim$30\% among SMGs with $L_{\rm IR} \sim 10^{13}$\,\lsun\ at $z \sim 3$. This estimate of the QSO duty cycle should be considered as an upper limit because we had assumed no BH accretion in SMGs without detectable QSOs. As the BH accretion rate ``flickers'' \citep[e.g.,][]{Hickox14}, an SMG may rise above and descend below the local relation multiple times throughout its lifetime of $\sim$200~Myrs \citep[e.g.,][]{Fu13}.

With average IR and AGN luminosities of ${\rm log}(L_{\rm IR}) \sim 13.1$~\lsun~$= 46.7$~erg~s$^{-1}$ and ${\rm log}(L_{\rm AGN}) \sim 46.6$~erg~s$^{-1}$, the mean IR luminosity of the SMG-QSO composite galaxies are $3-5\times$ greater than that of the general AGN population at $z \sim 2-3$ and the same bolometric luminosities \citep[e.g.,][]{Rosario12,Harris16}, confirming that the IR emission is dominated by unusually high star formation activity (consistent with the observation in Fig.~\ref{fig:sed}).

\section{Summary and Conclusions} \label{sec:summary}

We have observed 29 bright \hers\ 350~\um\ peakers within 30\arcsec\ of a QSO sightline with ALMA at 870~\um. The ALMA continuum images reach a resolution of $\sim$0.5\arcsec\ and an rms of $\sim$0.14\,mJy\,bm$^{-1}$, allowing us to measure the intrinsic sizes of the sources and to obtain precise astrometry for spectroscopic followup. Six of the \hers\ sources are less than 10\arcsec\ from the QSO, so that the QSOs are within the \hers\ beam as well as the ALMA primary beam. The \hers\ data indicate that these QSOs are likely associated with strong far-IR emission. The ALMA data allow us to check whether these are physical associations or line-of-sight projections. Our main findings are as follows:

\begin{enumerate}

\item We detect a total of 38 ALMA sources with flux densities between $0.7 \leq S_{870} \leq 14.4$~mJy within 10\arcsec\ of the \hers\ positions in 27 of the 29 ALMA fields. At 0.5\arcsec\ resolution, nine of the 27 \hers\ sources are resolved into multiple sources, and the remaining 18 \hers\ sources are single sources. No detection was made in two of the 29 fields, so these \hers\ sources may be spurious.

\item We confirm that 20 of the 29 \hers\ sources are SMGs with $S_{870} > 2$~mJy, and we identify 16 new SMG$-$QSO pairs that can be used to probe the CGM of SMGs.

\item ALMA robustly resolved 13 of the 38 sources. The sources are extremely compact: the mean beam-deconvolved 870~\um\ size is 0.29\arcsec$\pm$0.03\arcsec\ or 2.3\,kpc at $z = 3$.

\item The theoretical formula for positional uncertainty based on SNR and beam FWHM underestimates the true uncertainty -- almost half of our sample have \hers-ALMA offsets greater than the predicted values. Our empirical calibration suggests adding a baseline term of $\sim$0.7\arcsec\ to the formula to agree with the data (Eq.~\ref{eq:uncertainty}). 

\item Four out of the six \hers-detected QSOs are QSOs hosted by SMGs, or SMG-QSO composite galaxies. SED modeling indicates that the QSOs contribute a large fraction (30-60\%) of the total bolometric luminosity, although dust-obscured star formation dominates the far-IR emission. The BH accretion rates exceed the amount required to maintain the present-day BH-stellar mass ratio for their concurrent SFRs, and the exceeding amount suggests a luminous AGN fraction of $\lesssim$30\% in SMGs with $L_{\rm IR} \simeq 10^{13}$~\lsun. 

\item Two out of the six \hers-detected QSOs are close SMG-QSO pairs with separations at 8.9\arcsec\ and 11.9\arcsec. They are either line-of-sight projection or clustered sources. Based on the surface densities of \hers\ 350~\um-peakers and the number of high-redshift QSOs in the overlapping area, there should be $1.4\pm1.0$ projected pairs in a sample of six that have \hers-QSO separations between 5\arcsec\ and 10\arcsec. Thus at least one of the two is a projected pair with impact parameters less than 100~kpc. Probing the CGM of SMGs at such small impact parameters would be unprecedented. On the other hand, if the QSO and the SMG are in a merger, it would represent a rare case of ``wet-dry'' merger similar to SMM\,J04135$+$10277 \citep{Riechers13}. Spectroscopic redshifts are needed to separate between the two scenarios.

\end{enumerate}

The ALMA observations have provided sub-arcsec positions of 35 submillimeter sources near QSOs, enabling absorption line studies of the circumgalactic medium (CGM) of dusty starbursts at angular distances between 9\arcsec\ and 30\arcsec, or impact parameters between 71 and 240~kpc at $z \sim 3$. The four physical associations between luminous unobscured QSOs and bright \hers\ sources highlight an intense episode of the star-formation-AGN co-evolution over the cosmic history. However, their infiltration in samples of projected pairs poses a major challenge to our study of the CGM of dusty starburst galaxies. The limited spatial resolution of \hers\ is clearly the culprit. Even at an apparent \hers-QSO separation of 5$-$10\arcsec, two third of our ``pairs'' are QSOs hosted within SMGs. The fraction is likely to increase to essentially 100\% at \hers-QSO separations below 5\arcsec, making it difficult to probe the CGM of SMGs within $\sim$40~kpc. Resolving this issue relies on future far-IR and submillimeter observatories that enable wide-field extragalactic surveys at arcsec resolutions.

\acknowledgments
We thank K.~Gayley, J.~Hennawi, C.~Liang, R.~Mutel, and D.~Riechers for useful discussions. We also thank the anonymous referee for detailed comments that helped improve the presentation of the paper.
The National Radio Astronomy Observatory is a facility of the National Science Foundation (NSF) operated under cooperative agreement by Associated Universities, Inc. Support for this work was provided by the NSF through award GSSP SOSPA3-016 from the NRAO. 
H.F. acknowledges support from NASA JPL award RSA\#1568087, NSF grant AST-1614326, and funds from the University of Iowa. J.X.P. acknowledges support from the NSF grants AST-1010004, AST-1109452, AST-1109447, and AST-1412981.

This paper makes use of the following ALMA data: ADS/JAO.ALMA\#2015.1.00131.S. ALMA is a partnership of ESO (representing its member states), NSF (USA) and NINS (Japan), together with NRC (Canada), NSC and ASIAA (Taiwan), and KASI (Republic of Korea), in cooperation with the Republic of Chile. The Joint ALMA Observatory is operated by ESO, AUI/NRAO and NAOJ.
The \hers-ATLAS is a project with \hers, which is an ESA space
observatory with science instruments provided by European-led
Principal Investigator consortia and with important participation from
NASA. The H-ATLAS website is http://www.h-atlas.org/. The US
participants acknowledge support from the NASA \hers\ Science Center/JPL.
 
{\it Facilities}: ALMA, {\it Herschel}, Sloan, WISE

\bibliographystyle{/Users/fu/Documents/latex/apj/apj}
\bibliography{/Users/fu/Documents/bibliography/exgal_ref}

\clearpage

\clearpage

\appendix

\section{ALMA Images and Source Catalog}

\begin{figure*}[!tb]
\epsscale{1.15}
\plotone{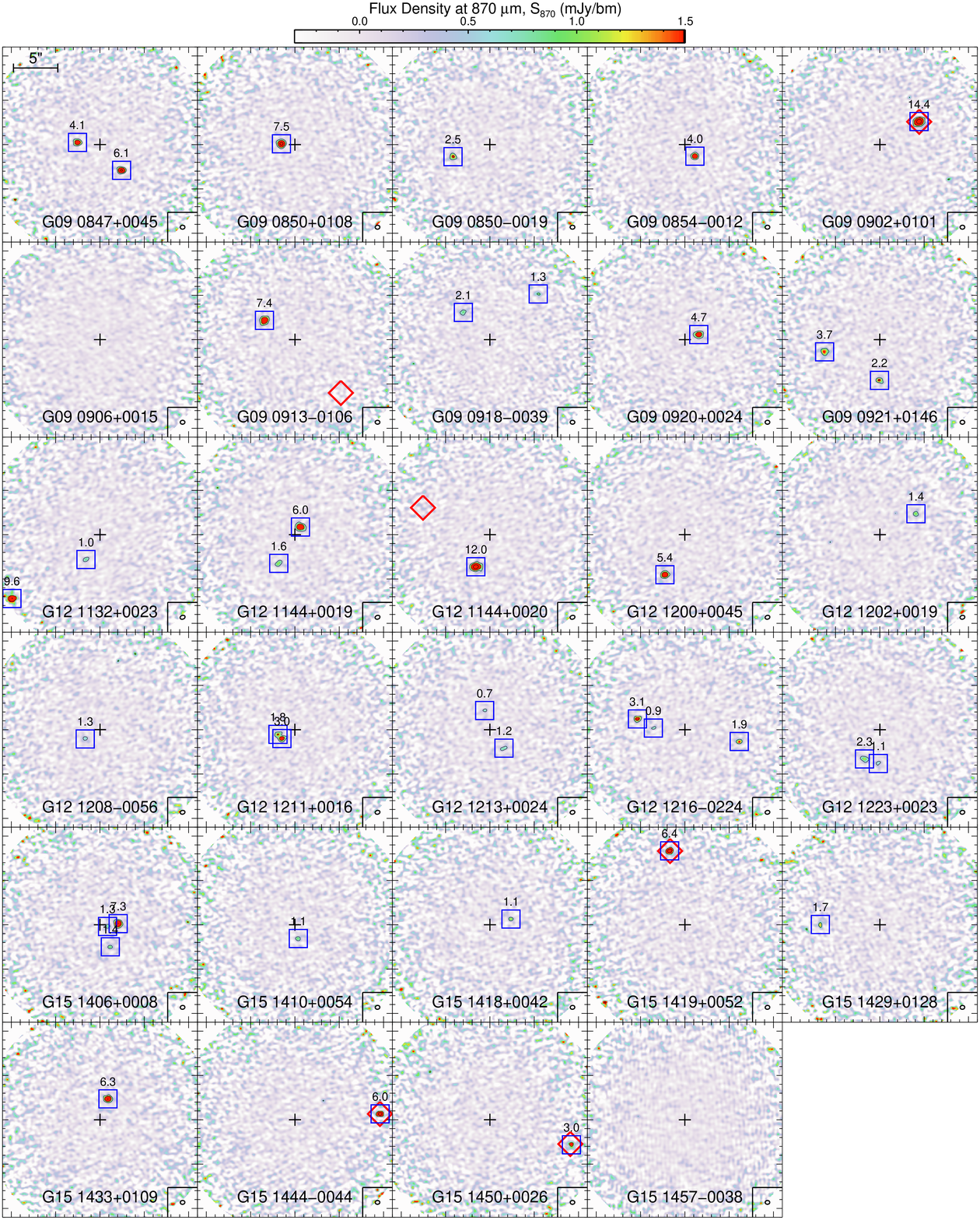}
\caption{ALMA 870~\um\ flux density maps of the entire sample. The contours are drawn at $(4,12,36)\times\sigma$. Detected sources are indicated by blue squares and their 870~\um\ flux densities are labeled in units of mJy. The position of the QSO is marked by a red diamond whenever it is within the field-of-view. 
\label{fig:alma_all}} 
\epsscale{1.0}
\end{figure*}

\begin{figure*}[!tb]
\epsscale{1.15}
\plotone{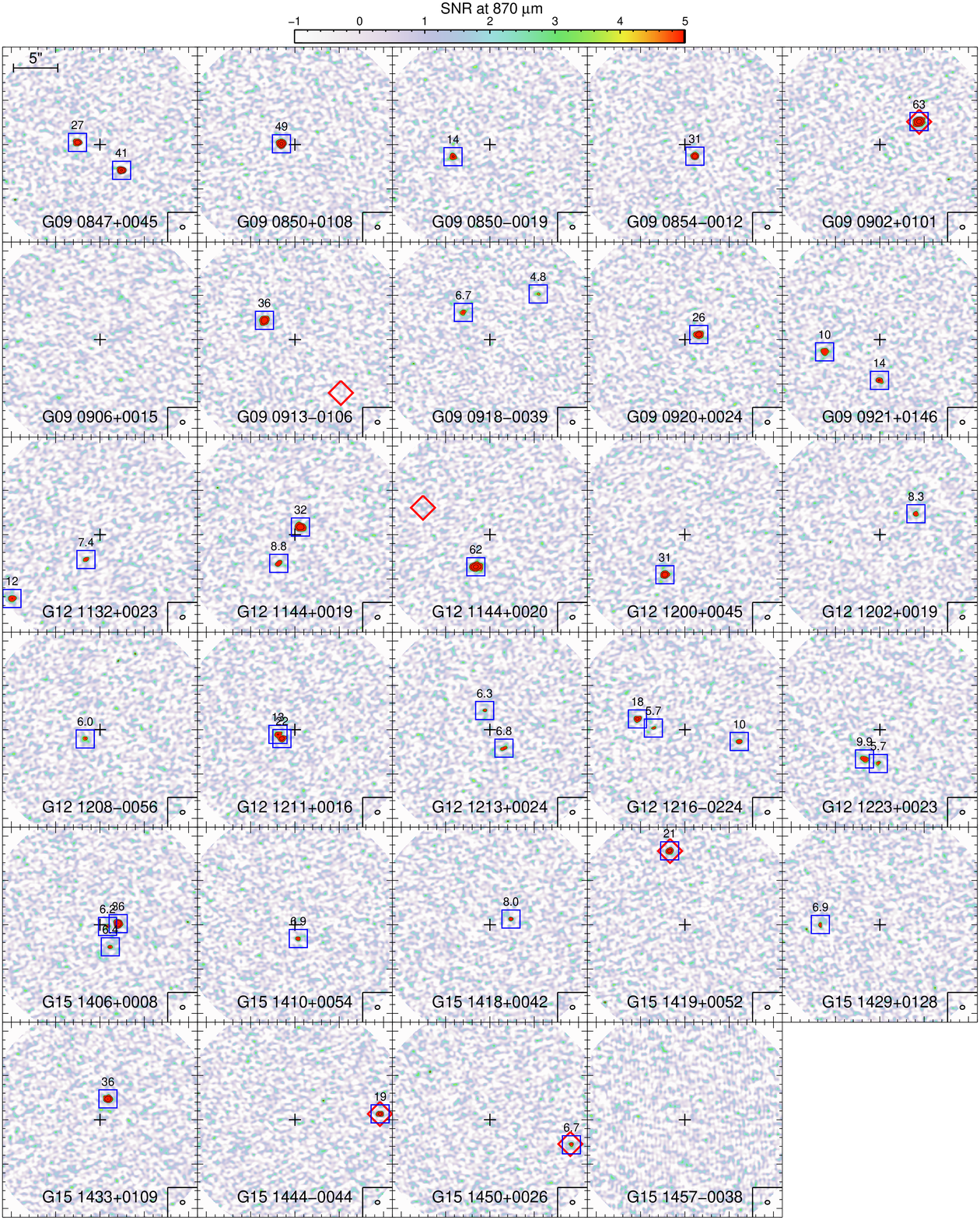}
\caption{Same as Fig.~\ref{fig:alma_all} but showing the 870~\um\ SNR maps. The peak SNR is labeled above each ALMA detection and is listed in Table~\ref{tab:alma}.
\label{fig:alma_snr}} 
\epsscale{1.0}
\end{figure*}

Figures~\ref{fig:alma_all} and \ref{fig:alma_snr} show, respectively, the ALMA flux density maps and the SNR maps for all of our 29 fields. Table~\ref{tab:alma} lists the source properties from \hers\ and ALMA. Out of the 29 \hers\ sources, 9 have multiple 870~\um\ counterparts, 18 have single 870~\um\ counterparts, and 2 are undetected. Note that the 9.6~mJy source in G12\,1132$+$0023 is not considered as the \hers\ source's counterpart because it is too far (12.3\arcsec) from the \hers\ 250~\um\ position. 

\begin{deluxetable*}{lccrrr ccrrrr}
\tablewidth{0pt}
\tablecaption{ALMA photometry and intrinsic source sizes.
\label{tab:alma}}
\tablehead{
\colhead{Field} & \colhead{R.A.$_{250}$} & \colhead{Decl.$_{250}$} & \colhead{$S_{250}$} & \colhead{$S_{350}$} & \colhead{$S_{500}$} & \colhead{R.A.$_{870}$} & \colhead{Decl.$_{870}$} & \colhead{offset} & \colhead{Peak S/N} & \colhead{$S_{870}$} & \colhead{Maj} \\
\colhead{} & \colhead{deg} & \colhead{deg} & \colhead{mJy} & \colhead{mJy} & \colhead{mJy} & \colhead{deg} & \colhead{deg} & \colhead{arcsec} & \colhead{} & \colhead{mJy} & \colhead{arcsec} \\
\colhead{(1)} & \colhead{(2)} & \colhead{(3)} & \colhead{(4)} & \colhead{(5)} & \colhead{(6)} & \colhead{(7)} &  \colhead{(8)} & \colhead{(9)} & \colhead{(10)} & \colhead{(11)} & \colhead{(12)}
}
\startdata
G09\,0847$+$0045&131.79600&+0.76550&34.2$\pm$6.4&47.5$\pm$7.8&38.8$\pm$8.4&131.79533&+0.76469& 3.8&41.5& 6.1$\pm$0.4&      $<$0.10\\
                &         &        &            &            &            &131.79671&+0.76557& 2.6&27.1& 4.1$\pm$0.3&0.25$\pm$0.07\\
G09\,0850$+$0108&132.72487&+1.13350&33.9$\pm$7.1&43.0$\pm$8.5&30.0$\pm$9.1&132.72529&+1.13352& 1.5&48.8& 7.5$\pm$0.5&0.32$\pm$0.03\\
G09\,0850$-$0019&132.72521&-0.32056&17.6$\pm$7.0&44.9$\pm$8.5&35.1$\pm$9.2&132.72636&-0.32094& 4.4&13.8& 2.5$\pm$0.3&0.39$\pm$0.12\\
G09\,0854$-$0012&133.55339&-0.20131&79.3$\pm$8.2&81.5$\pm$9.3&55.5$\pm$9.4&133.55308&-0.20167& 1.7&30.9& 4.0$\pm$0.3&      $<$0.17\\
G09\,0902$+$0101&135.65679&+1.02593&53.7$\pm$7.5&56.7$\pm$8.6&45.9$\pm$9.3&135.65556&+1.02665& 5.1&63.2&14.4$\pm$0.8&0.45$\pm$0.02\\
G09\,0913$-$0106&138.41382&-1.11628&52.5$\pm$7.4&69.4$\pm$8.8&48.4$\pm$9.2&138.41478&-1.11568& 4.1&35.8& 7.4$\pm$0.5&0.46$\pm$0.04\\
G09\,0918$-$0039&139.61586&-0.66437&36.9$\pm$7.2&48.1$\pm$8.5&29.4$\pm$9.2&139.61669&-0.66351& 4.3& 6.7& 2.1$\pm$0.5&0.63$\pm$0.22\\
                &         &        &            &            &            &139.61434&-0.66294& 7.5& 4.8& 1.3$\pm$0.4&      $<$0.65\\
G09\,0920$+$0024&140.24755&+0.40495&30.2$\pm$7.2&50.0$\pm$8.6&30.3$\pm$9.0&140.24711&+0.40511& 1.7&25.9& 4.7$\pm$0.4&0.45$\pm$0.05\\
G09\,0921$+$0146&140.26308&+1.77670&31.7$\pm$7.2&55.0$\pm$8.6&36.9$\pm$9.3&140.26308&+1.77542& 4.6&13.5& 2.2$\pm$0.3&0.38$\pm$0.10\\
                &         &        &            &            &            &140.26481&+1.77632& 6.4&10.3& 3.7$\pm$0.6&0.61$\pm$0.14\\
G12\,1132$+$0023&173.14797&+0.38582&22.1$\pm$7.0&39.3$\pm$8.5&25.7$\pm$9.1&173.15073&+0.38382&12.3&12.1& 9.6$\pm$1.4&0.51$\pm$0.16\\
                &         &        &            &            &            &173.14841&+0.38505& 3.2& 7.4& 1.0$\pm$0.2&      $<$0.54\\
G12\,1144$+$0019&176.15228&+0.33223&39.3$\pm$6.5&47.1$\pm$7.9&26.1$\pm$8.3&176.15211&+0.33247& 1.1&31.9& 6.0$\pm$0.4&0.53$\pm$0.05\\
                &         &        &            &            &            &176.15279&+0.33133& 3.7& 8.8& 1.6$\pm$0.3&0.48$\pm$0.17\\
G12\,1144$+$0020&176.06172&+0.34763&53.2$\pm$6.8&59.8$\pm$8.2&40.1$\pm$8.5&176.06216&+0.34663& 4.0&61.5&12.0$\pm$0.7&0.48$\pm$0.03\\
G12\,1200$+$0045&180.09801&+0.76231&49.7$\pm$7.5&57.9$\pm$8.7&39.5$\pm$9.1&180.09863&+0.76106& 5.0&30.6& 5.4$\pm$0.4&0.27$\pm$0.06\\
G12\,1202$+$0019&180.73549&+0.32368&22.9$\pm$6.9&43.7$\pm$8.4&28.4$\pm$8.9&180.73436&+0.32434& 4.7& 8.3& 1.4$\pm$0.3&      $<$0.53\\
G12\,1208$-$0056&182.06132&-0.94696&29.7$\pm$7.0&37.1$\pm$8.3&25.4$\pm$9.0&182.06178&-0.94725& 2.0& 6.0& 1.3$\pm$0.4&0.70$\pm$0.22\\
G12\,1211$+$0016&182.88821&+0.27572&26.8$\pm$6.3&35.4$\pm$7.7&34.3$\pm$8.3&182.88874&+0.27558& 2.0&12.8& 1.8$\pm$0.3&      $<$0.39\\
                &         &        &            &            &            &182.88862&+0.27545& 1.8&22.4& 3.0$\pm$0.3&0.26$\pm$0.08\\
G12\,1213$+$0024&183.34858&+0.41651&30.1$\pm$7.0&39.3$\pm$8.5&28.8$\pm$9.0&183.34814&+0.41593& 2.6& 6.8& 1.2$\pm$0.3&      $<$0.56\\
                &         &        &            &            &            &183.34874&+0.41710& 2.2& 6.3& 0.7$\pm$0.2&      $<$0.63\\
G12\,1216$-$0224&184.19284&-2.40837&48.2$\pm$7.4&53.2$\pm$8.6&27.7$\pm$8.9&184.19433&-2.40803& 5.5&17.9& 3.1$\pm$0.3&0.23$\pm$0.10\\
                &         &        &            &            &            &184.19114&-2.40874& 6.3&10.4& 1.9$\pm$0.3&0.33$\pm$0.14\\
                &         &        &            &            &            &184.19383&-2.40832& 3.6& 5.7& 0.9$\pm$0.3&      $<$0.63\\
G12\,1223$+$0023&185.91027&+0.38985&42.7$\pm$7.3&46.2$\pm$8.6&29.1$\pm$9.2&185.91075&+0.38893& 3.7& 9.9& 2.3$\pm$0.4&0.70$\pm$0.15\\
                &         &        &            &            &            &185.91032&+0.38879& 3.8& 5.7& 1.1$\pm$0.3&0.47$\pm$0.17\\
G15\,1406$+$0008&211.58725&+0.14376&72.4$\pm$8.0&79.7$\pm$9.1&44.5$\pm$9.3&211.58668&+0.14379& 2.1&36.3& 7.3$\pm$0.5&0.38$\pm$0.04\\
                &         &        &            &            &            &211.58692&+0.14307& 2.7& 6.4& 1.4$\pm$0.4&0.51$\pm$0.26\\
                &         &        &            &            &            &211.58701&+0.14370& 0.9& 6.2& 1.3$\pm$0.4&0.44$\pm$0.17\\
G15\,1410$+$0054&212.57149&+0.90727&38.5$\pm$7.2&46.7$\pm$8.5&28.7$\pm$9.0&212.57138&+0.90683& 1.6& 6.9& 1.1$\pm$0.3&0.25$\pm$0.19\\
G15\,1418$+$0042&214.57824&+0.70990&34.7$\pm$6.4&41.1$\pm$7.7&26.6$\pm$8.2&214.57758&+0.71008& 2.5& 8.0& 1.1$\pm$0.2&      $<$0.55\\
G15\,1419$+$0052&214.75198&+0.87667&44.3$\pm$6.8&49.3$\pm$8.1&31.5$\pm$8.7&214.75246&+0.87898& 8.5&20.8& 6.4$\pm$0.6&0.27$\pm$0.10\\
G15\,1429$+$0128&217.33184&+1.47928&26.0$\pm$6.3&38.4$\pm$7.7&35.8$\pm$8.4&217.33370&+1.47929& 6.7& 6.9& 1.7$\pm$0.5&      $<$0.60\\
G15\,1433$+$0109&218.33198&+1.15292&41.8$\pm$7.2&46.3$\pm$8.4&26.1$\pm$8.9&218.33173&+1.15358& 2.5&36.5& 6.3$\pm$0.4&0.29$\pm$0.05\\
G15\,1444$-$0044&221.10284&-0.74867&47.3$\pm$6.6&61.2$\pm$8.2&58.8$\pm$8.8&221.10017&-0.74849& 9.6&18.9& 6.0$\pm$0.6&      $<$0.28\\
G15\,1450$+$0026&222.67732&+0.43512&44.1$\pm$7.3&46.1$\pm$8.5&27.4$\pm$9.0&222.67477&+0.43434& 9.6& 6.7& 3.0$\pm$0.8&0.44$\pm$0.18
\enddata
\tablecomments{
 (1): Field designation. Note that in many cases ALMA resolved the \hers\ source into multiples.
 (2-3): \hers\ coordinates from the 250~\um\ detection.
 (4-6): \hers\ photometry at 250, 350, and 500~\um\ in mJy. The flux densities have been deboosted using the flux bias table of \citet{Valiante16}. The uncertainties include the 5.5\% systematic uncertainty for SPIRE absolute flux calibration. 
 (7-8): coordinates of all ALMA detections within 12.5 arcsec of the pointing position (i.e., the \hers\ 250~\um\ position).
 (9): angular offset from the \hers\ position in arcsec.
(10): 870~\um\ peak signal-to-noise ratio.
(11): integrated flux density at 870~\um\ in mJy based on Gaussian fits. The uncertainty includes the 5\% systematic uncertainty in the ALMA flux-density scale.
(12): Beam-deconvolved FWHM along the major axis. The ALMA source coordinates, the integrated flux densities, the beam-deconvolved FWHMs, and their associated uncertainties are all derived from elliptical Gaussian fits with the {\sc casa imfit} task.
}
\end{deluxetable*}

\end{document}